\documentclass{emulateapj}\usepackage{apjfonts}

\newcommand\hst{{\it HST}}
\newcommand\etal{{et~al.}} 
\newcommand\lta{\lesssim}
\newcommand\gta{\gtrsim}
\newcommand\ngc{\ensuremath{N_{\rm GC}}}

\newcommand\kms{km~s$^{-1}$}

\newcommand\Iacs{\ensuremath{I_{814}}}

\newcommand\mM{\ensuremath{(m{-}M)}}
\newcommand\mo{$M_{\odot}$}
\newcommand\M{\ensuremath{\mathcal M}}
\newcommand\bg{\hbox{\textit{bg}}}
\newcommand\smass{\ensuremath{S_\mathcal{M}}}
\newcommand\sharpdao{\textit{sharp}$_{\rm\,DAO}$}
\shorttitle{Globular Clusters in A1689}
\shortauthors{Alamo-Mart\'inez et al.}

\begin{document}

\title{The Rich Globular Cluster System of Abell 1689 and the Radial Dependence of the Globular Cluster Formation Efficiency}

\author{K. A. Alamo-Mart\'{i}nez\altaffilmark{1,2},
  J. P. Blakeslee\altaffilmark{2}, M. J. Jee\altaffilmark{3},
  P.~C\^ot\'e\altaffilmark{2}, L.~Ferrarese\altaffilmark{2},
  R.~A.~Gonz\'alez-L\'opezlira\altaffilmark{1},
  A.~Jord\'an\altaffilmark{4}, G.~R. Meurer\altaffilmark{5},
  E.~W.~Peng\altaffilmark{6}, M.~J.~West\altaffilmark{7,8}}

\altaffiltext{1}{Centro de Radioastronom\'ia y Astrof\'isica, Universidad Nacional Aut\'onoma de M\'exico, Morelia 58090, M\'exico; {k.alamo@crya.unam.mx}}
\altaffiltext{2}{Herzberg Institute of Astrophysics, National Research Council of Canada, Victoria, BC V9E 2E7, Canada}
\altaffiltext{3}{Department of Physics, University of California, Davis, One Shields Avenue, Davis, CA 95616, USA}
\altaffiltext{4}{Departamento de Astronom\'ia y Astrof\'isica, Pontificia Universidad Cat\'olica de Chile, 7820436 Macul, Santiago, Chile}
\altaffiltext{5}{International Centre for Radio Astronomy Research, The University of Western Australia, 35 Stirling Highway, Crawley, WA 6009, Australia}
\altaffiltext{6}{Department of Astronomy \&
 Kavli Institute for Astronomy and Astrophysics, Peking University, Beijing 100871, China}
\altaffiltext{7}{European Southern Observatory, Alonso de C\'ordova 3107,
  Vitacura, Santiago, Chile} 
\altaffiltext{8}{Maria Mitchell Observatory, 4 Vestal Street, Nantucket, MA 02554, USA}

\begin{abstract}
We study the rich globular cluster (GC) system in the center of the massive cluster of
galaxies Abell~1689 ($z=0.18$), one of the most powerful gravitational lenses known.
With 28 {\sl HST}/ACS orbits in the F814W bandpass, we reach
magnitude $I_{814}=29$\ with $\gtrsim\,$90\% completeness and sample the brightest
$\sim\,$5\% of the GC system.
Assuming the well-known Gaussian form of the GC luminosity function (GCLF), we
estimate a total population of 
$N^{\rm total}_{\rm GC} = 162,850^{+75,450}_{-51,310}$ GCs within a
projected radius of 400~kpc.  As many as half may comprise
an intracluster component.  Even with the sizable uncertainties, which mainly result from
the uncertain GCLF parameters, this is by far the largest GC system studied to date.
The specific frequency $S_N$ is high, but not uncommon for central galaxies
in massive clusters, rising from $S_N\approx5$ near the center to $\sim12$ at
large radii. Passive galaxy fading would increase $S_N$ by $\sim20\%$ at $z{\,=\,}0$. 
We construct the radial mass profiles of the GCs, stars, intracluster gas, and
lensing-derived total mass, and we compare the mass fractions as a function of radius.  
The estimated mass in GCs, $\M_{\rm GC}^{\rm total}$=3.9$\times{10}^{10}$\mo, 
is comparable to $\sim$80\% of the total stellar mass of the Milky Way.
The shape of the GC mass profile appears intermediate between those of the 
stellar light and total cluster mass.  Despite the extreme nature of this system,
the ratios of the GC mass to the baryonic and total masses, and thus the GC
formation efficiency, are typical of those in other rich clusters when comparing
at the same physical radii.  The GC formation efficiency is not constant,
but varies with radius, in a manner that appears similar for different clusters;
we speculate on the reasons for this similarity in profile.
\end{abstract}
\keywords{globular clusters: general --- galaxy clusters: individual(Abell 1689)}

\section{Introduction}

In the usual hierarchical structure formation model, pregalactic objects
begin forming through gravitational instability in high-density regions and
coalesce to form progressively larger structures (e.g., Springel et al.\ 2005;
De Lucia et al.\ 2006).
Massive clusters of galaxies, the largest self-gravitating systems in the universe
today, saw the earliest major star formation and have been assembling for the
longest time.  The rapid merging and accretion in these dense
environments has obscured much information about the early dynamical histories.
However, high-resolution numerical simulations indicate
that old metal-poor globular clusters (GCs) and diffuse stellar light
provide reliable tracers of the earliest star-forming substructures (Moore et
al.\ 2006; Abadi et al.\ 2006), including their spatial distributions and
kinematics.  In contrast, the surviving dwarf satellites likely originated
at lower densities and were accreted later 
(see Font et al.\ 2006; De Lucia et al.\ 2008; Johnston et al.\ 2008).

GCs are especially useful tracers because they are abundant in all large
galaxies and most dwarfs with masses of at least $10^8 M_\odot$ (e.g., Peng et
al.\ 2008).  As in the Milky Way, they generally have ages $\gtrsim10$~Gyr and
peak metallicities just a few percent of solar (e.g., Cohen \etal\ 1998, 2003;
Puzia \etal\ 2005; Chies-Santos et al.\ 2011, 2012), consistent with their being
relicts of early accretion.  Harris \& van den Berg (1981) introduced the
specific frequency $S_N$ (the number of GCs per unit galaxy luminosity) as a
normalized measure of population richness. 
$S_N$ has been measured for galaxies of all morphological types and environments
(see reviews by Harris 1991, 2001; West et al.\ 2004; Brodie \& Strader 2006), 
though most commonly in early-type galaxies, as spirals generally have much sparser GC systems.
For early-type dwarfs, $S_N$ exhibits a large scatter, with values ranging from
zero to $\sim\,$100, and some tendency to increase as the luminosity decreases. 
Conversely, for more luminous ellipticals, 
$S_N$ ranges from $\sim\,$2 to $\sim\,$10, and tends to
increase with luminosity. 

Similar trends of $S_N$ with luminosity occur for early-type galaxies in both
clusters and isolated environments (e.g., Cho et al.\ 2012; Alamo-Mart{\'i}nez
et al.\ 2012).  The exception is for galaxies near the dynamical centers of
massive clusters, where the total GC population often exceeds 10,000, and the
central cD galaxy may have $S_N>10$ (Harris et al.\ 1995; West et al.\ 1995).
Blakeslee et al.\ (1997) found that the number \ngc\ of GCs in such systems
scales approximately with 
cluster velocity dispersion, X-ray luminosity, and other indicators of cluster
mass, rather than with host galaxy luminosity.  Thus, the number per unit total 
(halo) mass shows much less variation than $S_N$.
McLaughlin (1999) found a similar scaling of \ngc\ in massive ellipticals with the
baryonic mass; Blakeslee (1999) showed that the ratio of the scale factors was
consistent with the expected baryon fraction.
Numerical models also predict an approximately constant frequency of old metal-poor
GCs relative to total halo mass, depending on the degree of local variation in the 
epoch of reionization (Moore \etal\ 2006).


Peng et al.\ (2008) studied $S_N$ and the stellar mass fraction contained in GCs
for 100 early-type galaxies from the Advanced Camera for Surveys (ACS) Virgo
Cluster Survey (C\^ot\'e et al.\ 2004).
They found that the GC stellar mass fraction tends to be larger in giant and
dwarf galaxies, but is universally low at intermediate masses. 
Such intermediate mass galaxies also appear to have been most efficient
in converting baryons into stars
(e.g., van den Bosch et al.\ 2007; Conroy \& Wechsler 2009; Guo et al.\ 2010). 
Thus, as previously found for central cluster galaxies, 
the $S_N$ and GC stellar mass fraction
for the ensemble of Virgo galaxies scale more closely
with halo mass (mainly dark matter) than with the stellar mass or luminosity.
The observed variations therefore indicate a variable field star formation efficiency
following the GC formation epoch. 
Spitler \& Forbes (2009) compiled a sample of galaxies with a wide 
range of masses in various environments and also concluded that there was
a direct proportionality between the mass in GCs and the total halo mass.
Following this idea, Georgiev et~al.\ (2010) studied GC formation efficiencies within
an analytical model that included mass-dependent feedback
mechanisms.  They found that the GC formation efficiency is roughly constant with halo
mass, but $S_N$ and stellar mass fraction 
vary with the field star formation efficiency, which 
in turn depends on halo mass and is highest at intermediate values.
In this model, the increased scatter among dwarf galaxies was attributed to
stochastic effects at low mass.
However, Peng et~al.\ had found that the Virgo dwarfs with high GC mass
fractions were nearly all within 1~Mpc of M87, indicating that location plays a 
key role in determining the relative fractions of baryons in GCs and field
stars in these systems.  This could also be a consequence of the formation
epoch, with GC formation efficiency being higher at early times (e.g., Kruijssen 2012).


In the Coma cluster, Peng et al.\ (2011) found a very large population of
intracluster globular clusters (IGCs), which are bound to the cluster 
potential, rather than to individual galaxies. Limits on the amount of 
diffuse intracluster light (ICL) imply a high $S_N$ for the IGC population, 
similar to the values for high-$S_N$ cD galaxies and the centrally located 
Virgo dwarfs. 
IGCs appear to be a common feature of galaxy clusters: a significant population 
likely resides in Abell~1185 (Jord\'an et al.\ 2003; West et~al.\ 2011), 
a cluster in which the cD galaxy is offset from the centroid of X-ray emission;
Lee et al.\ (2010) also report evidence for IGCs in Virgo.  
The IGC populations are overwhelmingly metal-poor, with colors typical of GCs in dwarf
galaxies and the outer regions of massive galaxies.  Because of their early formation
and subsequent dissipationless assembly, the spatial density profile of the IGCs is
expected to trace the total cluster mass profile.  However, because of their low
surface densities and contamination from GCs bound to galaxies, it remains unclear
whether they follow more closely the stellar light,
the baryonic matter (including the X-ray gas), or the dark matter distribution.  
Observations thus far have
been limited to nearby ($z\lta0.05$) moderate mass clusters.  Further progress
requires studying massive clusters with many thousands of GCs and well characterized
baryonic and total mass density profiles.

Abell~1689 (hereafter A1689) is an extremely massive galaxy cluster at $z=0.183$, with a
velocity dispersion $\sigma\gtrsim1400$ \kms, and complex kinematical
substructure (e.g., Teague et al.\ 1990; Girardi et al.\ 1997; Lemze et
al.\ 2009).  It has one of the largest known Einstein radii, and its mass
distribution has been extensively studied from both strong and weak lensing
(Tyson \& Fischer 1995; Taylor et al.\ 1998; Broadhurst et al.\ 2005, Zekser et
al.\ 2006; Limousin et al.\ 2007; Coe et al.\ 2010), as well as X-ray properties
(Andersson \& Madejski 2004; Lemze et al.\ 2008).
A recent multi-wavelength analysis by Sereno et al.\ (2013), including
Sunyaev-Zeldovich, X-ray, and lensing data, finds a total mass
$M_{200} = (1.3\pm0.2)\times10^{15} M_\odot$ within $r_{200} = 2.1$~Mpc.

A1689 was one of the targets selected for early release observations with the
ACS Wide Field Channel (ACS/WFC; Ford et al.\ 2002, 2003)
after installation on the \textit{Hubble Space Telescope} (\hst). 
In addition to the strong lensing analyses referenced above, the same observations 
were used to investigate ultra-compact dwarf galaxies in an extremely 
rich environment (Mieske et al.\ 2004). 
Further inspection of these
early ACS data revealed a concentration of faint point sources that were consistent with GCs
at the distance of A1689.  Assuming the usual Gaussian GC luminosity function (GCLF)
implied a huge population of $\gtrsim10^5$~GCs (Blakeslee 2005),
but the number was very uncertain,
as it involved an extrapolation by two orders of magnitude.  

Here we report results from very deep \hst/ACS observations of A1689 obtained in
order to test the previous uncertain results and study the GC population in the
center of this extraordinary cluster.  The following section summarizes the
observations and image reductions.  
Section~\ref{analysis.sect} describes the
photometric analysis of the galaxies and GCs in detail, while
Section~\ref{N_Sn.sect} presents our results on the GC number density
distribution, total population, and specific frequency.  
In Section~\ref{results.sect}, we compare our results on the GC and galaxy light
distributions with the mass profiles of the X-ray emitting gas and dark matter.
The final section discusses our conclusions.  Throughout this work, we adopt the
WMAP7 maximum likelihood cosmology (Komatsu et al.\ 2011) with
$(h,\Omega_m,\Omega_{\lambda}) = (0.704,0.27,0.73)$, which yields
a distance modulus for A1689 of $\mM = 39.74$ mag and a physical scale of
3.07~kpc~arcsec$^{-1}$ (These numbers change by $<0.5$\% for the WMAP9 cosmology
of Hinshaw et al.\ 2013).

\section{Observations and Image Reduction}
\label{obs.sect}

\begin{figure}
\centering
\includegraphics[angle=0,width=0.45\textwidth]{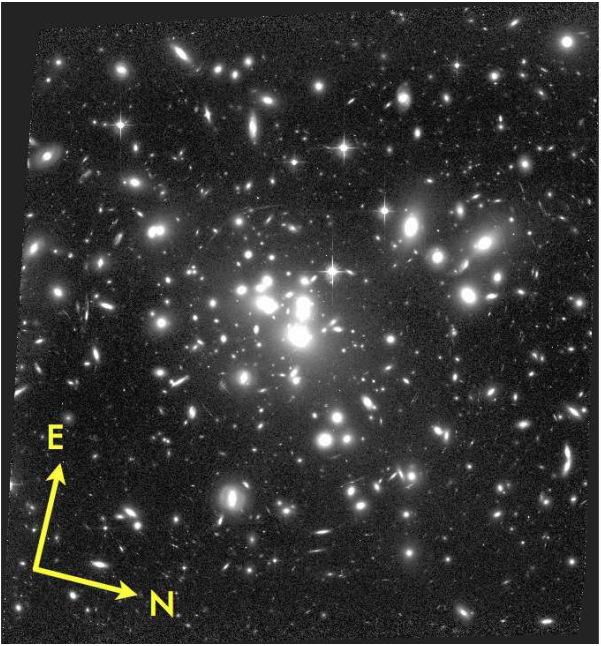}
\caption{Deep ACS/WFC F814W image of the galaxy cluster A1689 from Program GO-11710,
shown in the observed orientation. The field-of-view is $\sim$3\farcm3$\times$3\farcm3. 
 \label{A1689_original}}
\end{figure}

As part of \hst\ Program GO-11710, we imaged the central field of A1689 for 28~orbits
with the $F814W$ bandpass (also referred to as \Iacs) of the ACS/WFC during seven visits
from 2010 May~29 to 2010 July~08.  The charge transfer efficiency (CTE) of the
ACS/WFC detectors has become significantly degraded in recent years.  After
standard Space Telescope Science Institute pipeline processing to the point of
calibrated ``flt~files,'' we therefore used the stand-alone script version of the
empirical pixel-based CTE correction algorithm of Anderson \& Bedin (2010) on each of
the 56 individual exposures in our program.  This correction script did an excellent job of
removing the CTE trails in the data.  

The individual exposures were then processed with Apsis (Blakeslee et al.\ 2003) to
produce a single geometrically corrected, cosmic-ray cleaned, stacked image with a
total exposure time of 75,172\,s.  At nearly 21~hr, this is the single deepest
ACS/WFC image in the F814W bandpass.  After experimenting with different Drizzle
(Fruchter \& Hook 2002) parameters within Apsis, we adopted the Gaussian
interpolation kernel with a ``pixfrac'' of 0.5 and an output pixel scale 
of 0\farcs033~pix$^{-1}$.  This set of parameters provides improved 
resolution with good sampling of the point spread function (PSF), 
without introducing excessive small-scale correlations in the pixel noise.

We calibrated the photometry on the AB system, using the F814W zero~point of
25.947 determined from the time-dependent 
ACS Zeropoint Calculator.\footnote{http://www.stsci.edu/hst/acs/analysis/zeropoints.}
The \Iacs-band Galactic extinction in the direction of
A1689 is 0.04~mag (Schlafly \& Finkbeiner 2011).


\begin{figure*}
\centering
\includegraphics[angle=0,width=0.95\textwidth]{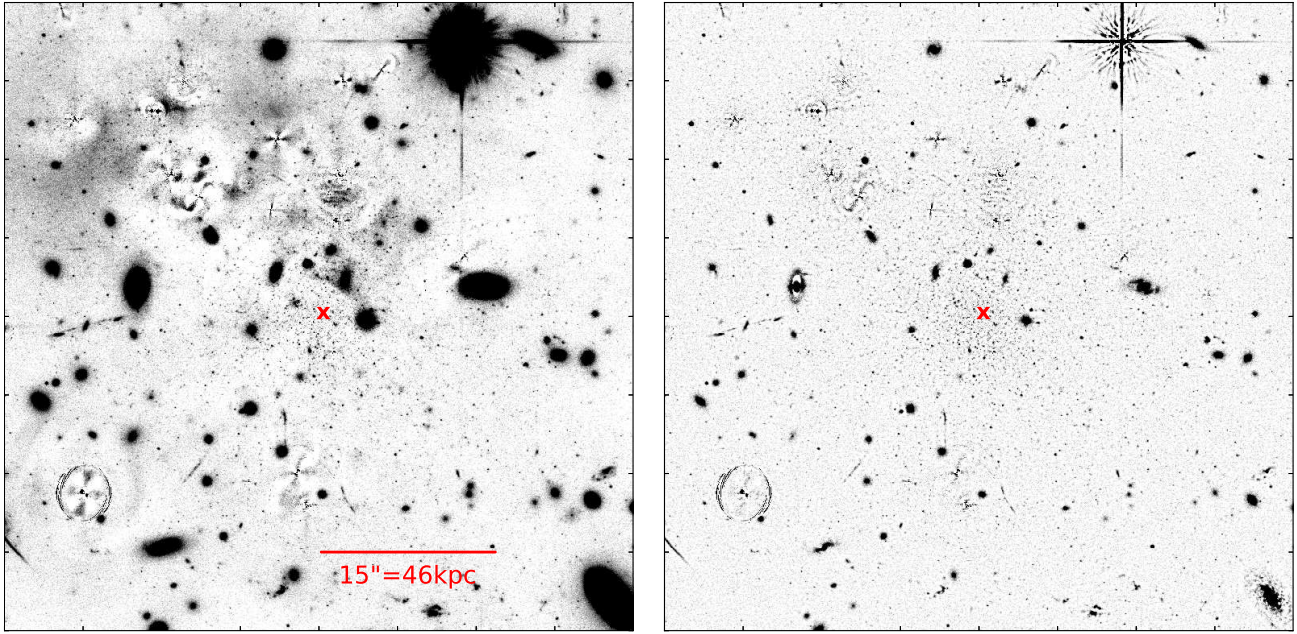}
\caption{Zoom to the central region ($\sim160{\times}160$ kpc) of A1689. The red 
cross marks the center of the cD galaxy.
{\it Left panel:} residual after subtraction of the luminosity model (\textit{bmodel} plus
SExtractor background map); because of its shallow light
profile, the cD galaxy itself subtracts very well. 
{\it Right panel:} residual after subtraction of the luminosity model and the
smooth rmedian image, used only for object detection. \label{Zoom_rmed}} 
\end{figure*}

\section{Analysis}
\label{analysis.sect}

\subsection{Galaxy Modeling} 
\label{galaxymodel.sect}

As one of the few richness class~4 clusters in the Abell catalogue (Abell et al.\ 1989),
A1689 is exceptionally rich in galaxies, particularly in its central region 
(Figure~\ref{A1689_original}),
a fact that hampers source detection and photometry.
In order to have an image with the flattest background possible, we used
the {\sl ellipse} (Jedrzejewski 1987) and {\sl bmodel} tasks within IRAF \citep{tody86,tody93}
to construct isophotal models for 59 of the
brightest galaxies in the ACS/WFC field.  Neighboring galaxies were masked 
during the fitting, and in cases with close companions, it was necessary 
to perform several iterations.

The isophotal models for the 59 galaxies were combined to produce a single \textit{bmodel} image,
which was then subtracted from 
the original image to create a first-pass residual image. SExtractor \citep{bert96} 
was then run on this residual image to generate a map of 
the background due to the imperfect subtraction of the galaxies, as well as approximate 
representations of other, unmodeled, cluster galaxies.
The SExtractor background map was then combined with the \textit{ellipse/bmodel} models to produce our 
final luminosity model, and this was subtracted from the original image to obtain
what we refer to as the ``final\_residual'' image. 

\subsection{Object Detection} 
\label{objectdetection.sect}

The final\_residual image includes many smaller galaxies and residuals from larger galaxies
that were difficult to model and not well represented by the SExtractor background map. 
To remove these smaller (but resolved) structures, 
a very smooth image was created with the IRAF task {\sl rmedian} 
(a ring median filter with inner and outer radii of~5 and 9~pix, respectively)
applied to the final\_residual image.
This ring-median image was then subtracted from the final\_residual
to obtain the  ``rmed\_residual'' image (Figure~\ref{Zoom_rmed}). Although the rmed\_residual 
image is extremely flat and suitable for point source detection, 
it cannot be used to measure reliable source~magnitudes, 
as some flux is removed by the rmedian process.

The source detection was performed with SExtractor on the rmed\_residual image.
In order to make the detection more robust, we used an RMS error image for the
SExtractor detection WEIGHT map, produced as described by Jord\'an et~al.\ (2004). 
This RMS map includes detector and photometric noise, as well as
the signal-to-noise variations from the corrected bad pixels and cosmic rays.
Bright stars, diffraction spikes, areas of lower exposure time near the image
edges, and regions with large model residuals (due to sharp or irregular features
within the cluster galaxies) were masked during the detection 
(black regions in Figure~\ref{fig_xy_nGC}, right panel).
At a luminosity distance of 885~Mpc, the GCs appear as point sources, and the SExtractor 
parameters were chosen to optimize point source detection, using a threshold of 
5~or~more connected pixels at least 1.5~$\sigma$ above the local background.

Source magnitudes were measured on the final\_residual image with PSF photometry, 
using the SExtractor output coordinates. The PSF photometry is more accurate 
than the various aperture magnitudes measured by SExtractor.
The PSF was constructed using the standard 
DAOPHOT \citep{stet87} procedure. The important parameters were 
the FWHM of the PSF fwhmpsf\,=\,2.6 pix (0\farcs086), aperture\,=\,4 pix, and varorder\,=\,2,
which means that the PSF is quadratically variable over the image.
The best fitting function to describe the PSF was {\sl penny1}.\footnote{An elliptical Gaussian core, 
that can be tilted at an arbitrary angle, with Lorentzian wings.} 
Since the radius of the PSF model is finite, an aperture correction, 
$m_{\rm apcor}$, was estimated. To this end, first the magnitude difference between 
4 and 15 pixels ($\sim$0$\farcs$5) was measured, then a correction from an
aperture of 0$\farcs$5 to infinity was obtained from \citet{siri05}.  The final
aperture correction $m_{\rm apcor}=-0.36$~mag was applied to all the measured 
DAOPHOT fit magnitudes.

\subsection{GC Candidate Selection} 
\label{gcselection.sect}

\begin{figure*}[!h]
\begin{center}
$\vcenter{\hbox{\includegraphics[angle=0,width=0.5\textwidth]{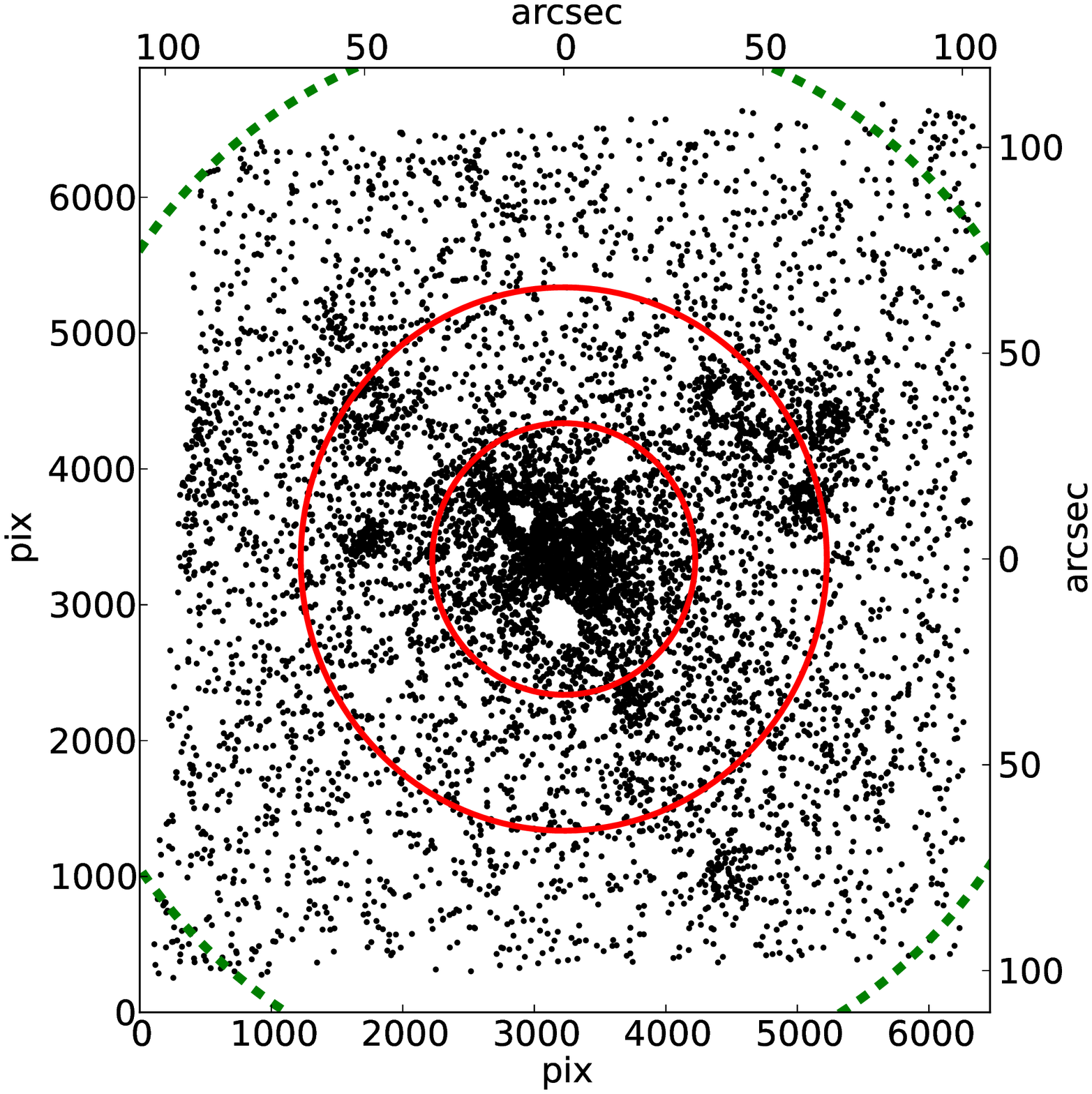}}}$
$\vcenter{\hbox{\includegraphics[angle=0,width=0.45\textwidth]{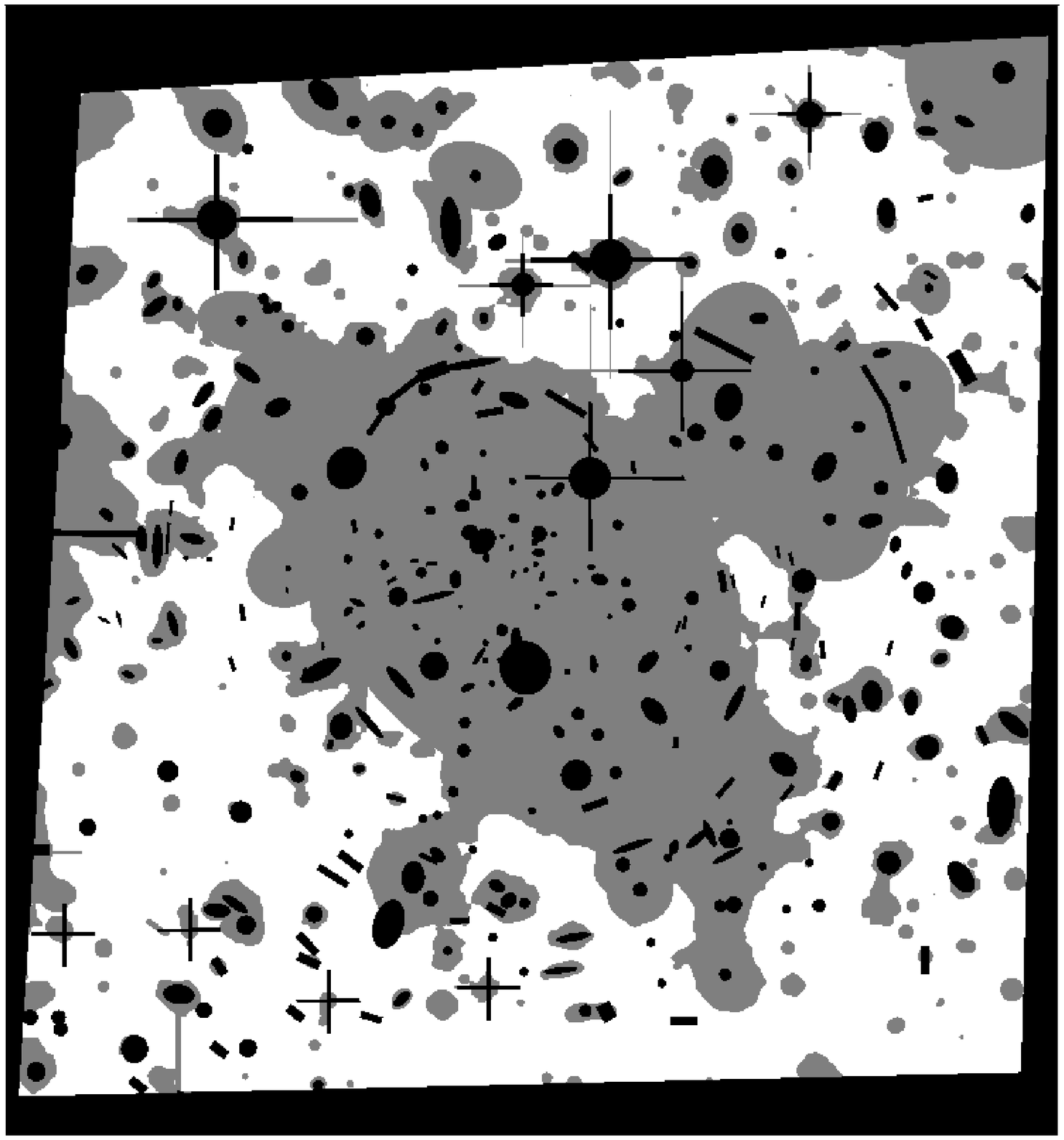}}}$
\end{center}
\caption{{\it Left panel:} spatial distribution of GC candidates; the red circles
  mark the boundaries of the three separate regions where the completeness function
  was fitted, and the green dashed circle indicates a radius of 400 kpc (130\arcsec). 
{\it Right panel}: black regions indicate the areas masked throughout the entire
analysis; the gray regions show the masks applied for the galaxies, including the cD,
used in estimating the number of background contaminants, \bg\ (see Sect.\,\ref{bgcont.sect}).  
\label{fig_xy_nGC}}
\end{figure*}

The DAOPHOT parameters $\chi_{\rm DAO}$ and \sharpdao, which gauge the 
goodness of the PSF fit and profile sharpness for each object,
together provide a good indication of whether an object is a point source.
Based on input and output parameters of artificial stars constructed 
from the PSF, we selected point sources as objects with 
$\chi_{\rm DAO}<$5 and $-0.9<\hbox{\textit{sharp}}_{\rm DAO}<0.9$.

Assuming that they are similar to nearby globular clusters, the GCs in A1689 should 
appear at $\Iacs\gta27.0$~AB.  We therefore selected GC candidates as point sources
having $27.0<\Iacs<29.32$; the faint limit is the magnitude where our detection is  
50\% complete in the innermost region of the cluster (see below). 
From this combined selection, we obtained a sample of 8212 GC candidates;
Figure~\ref{fig_xy_nGC} shows their $x,y$ positions, which are strongly
concentrated near the cD galaxy, but excesses of GC candidates are also
visible around other cluster galaxies.

\subsection{Completeness} 
\label{completeness.sect}

\begin{figure*}
\centering
\epsscale{1}
\plottwo{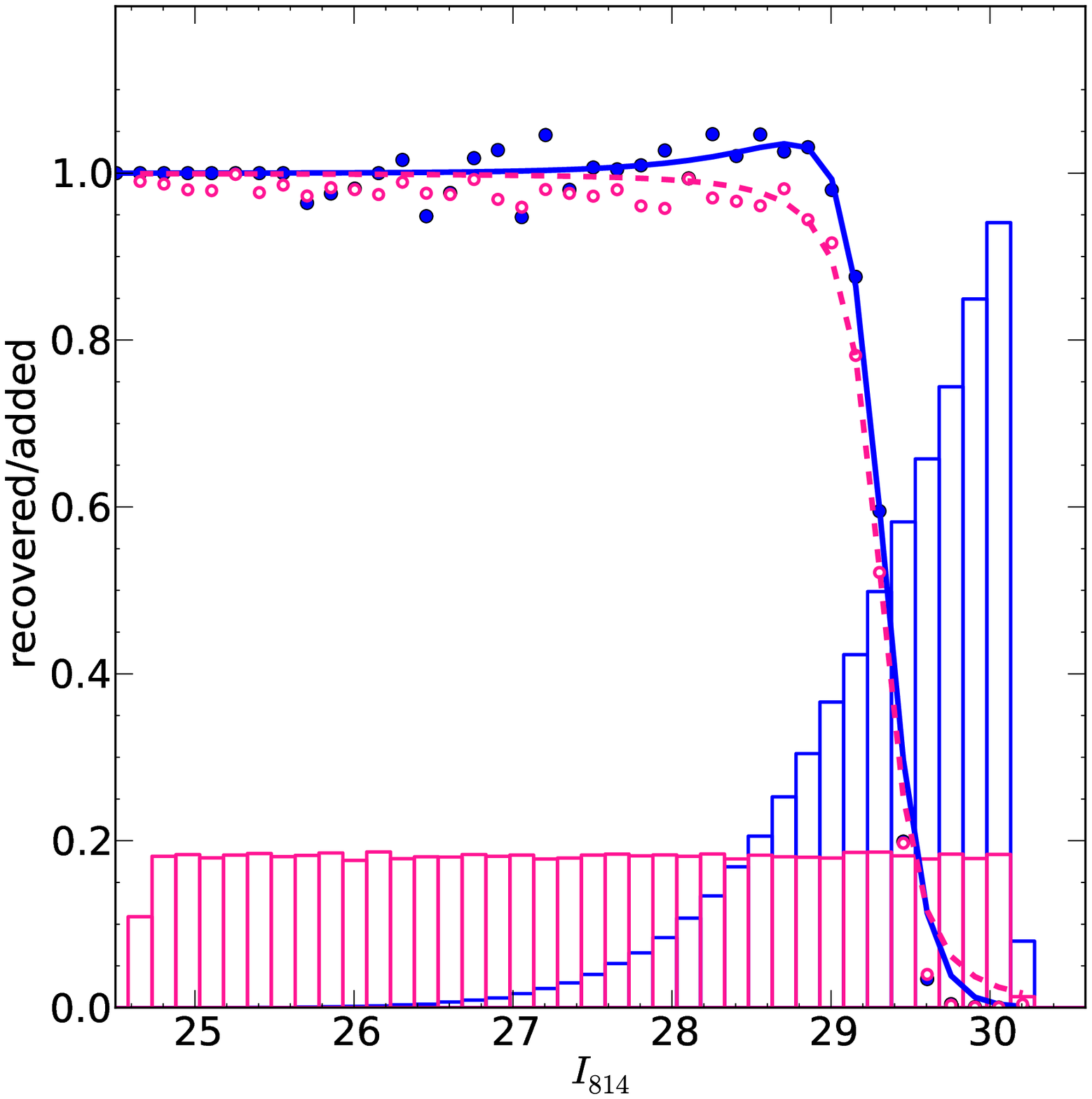}{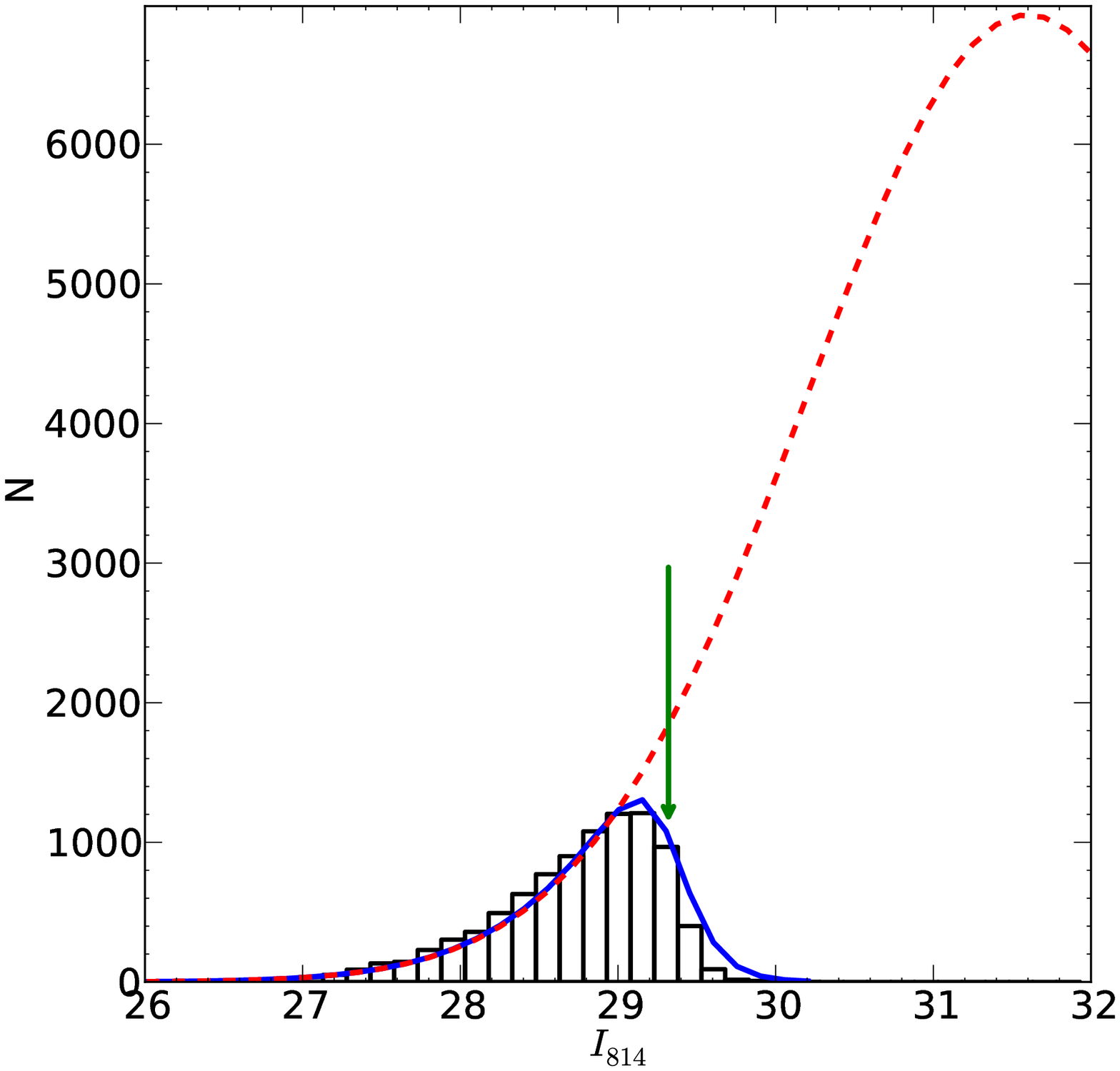}
\caption{{\it Left panel:} completeness functions for the two different input magnitude distributions.
{\it Histograms:} magnitude distributions of the artificial stars, i.e., a Gaussian with 
mean $\mu=$31.6 mag and $\sigma=$1.5 mag ({\it blue}), and uniform ({\it pink}). 
The blue points and pink open circles are the respective recovered fractions,
and the lines show the respective best-fit completeness functions: Pritchet (dashed
pink) and modified Fermi (solid blue). 	
The rising Gaussian distribution, meant to mimic the actual GCLF, causes an excess of 
recovered point sources near the completeness limit because of Eddington bias.
{\it Right panel:} observed luminosity function of GC candidates (histogram); 
fitted modified Fermi function $\times$ Gaussian model (solid blue line);
and the derived Gaussian GCLF (dashed red line). The green arrow at $\Iacs=29.32$ 
indicates the magnitude limit used for the fits. \label{fig_completeness}}
\end{figure*}

To quantify the completeness, 250\,000 artificial stars  were constructed from the 
PSF model and added 500 at a time with random $r,\theta$ positions. The origin of the 
polar coordinates was the center of the cD galaxy, and the uniform random distribution in $r$ 
yielded a higher density of sources near the cluster center, mimicking the actual sources.
In adding the artificial sources, the masked areas were avoided, 
and the added sources were not allowed to overlap with each other.
The artificial stars were added to the rmed\_residual and final\_residual images, 
and their fluxes measured with the same procedure that was followed 
for the real objects, including the selection based on the values of 
$\chi_{\rm DAO}$ and \sharpdao. 
We carried out this whole process twice, using two different magnitude distributions
for the artificial sources:
(1) a uniform, or box-shaped, distribution with $25.0<m<30.5$; 
(2) a Gaussian distribution with mean $\mu=$31.6, $\sigma=$1.5, and 
the constraint $m<30.5$ (more than a magnitude beyond the completeness limit).
The latter case approximates the expected GCLF in A1689 (see Sec.~\ref{N_Sn.sect});
both distributions are illustrated in Figure~\ref{fig_completeness}.

In the case of the uniform magnitude distribution, the fraction of recovered stars as a 
function of magnitude is well described by a function of the form:
\begin{equation}
f^P(m)=\frac{1}{2}\left[1-\frac{\alpha(m-m_{\rm lim})}{\sqrt{1+\alpha^2(m-m_{\rm lim})^2}}\right]\,,
\label{eq.pritchet}
\end{equation}
where $m_{\rm lim}$ is the magnitude at which the completeness is 0.5, and $\alpha$ 
determines the steepness of the curve.  This function $f^P$
is sometimes referred to as a ``Pritchet function'' (e.g., McLaughlin et al.\ 1994).

In the case of the Gaussian magnitude distribution, the fraction of recovered
stars actually exceeds unity before the steep drop in completeness sets in.
This excess of detected sources is due to Eddington (1913) bias:
as a result of the steeply rising luminosity function, measurement errors cause more
faint sources to be scattered to brighter detection magnitudes, and relatively 
fewer bright sources to be scattered to fainter levels.
In this case, including both magnitude bias and incompleteness, the recovered 
fraction was not well described by Eq.~(\ref{eq.pritchet}).  However, we found 
that it could be represented by the following modified version of the Fermi function:
\begin{equation}
f^F(m)=\frac{1+C \exp{[b(m-m_0)]}}{1+\exp{[a(m-m_0)]}}\,,
\end{equation}
where $m_0$ is the magnitude at which the completeness would be 0.5 for 
a standard Fermi function ($C\equiv0$).  
The other parameters are linked, but in rough terms,
$a$ controls the steepness of the cutoff, 
$b$ (which must be $<a$) affects where the departure above unity begins,
and $C$ determines the amplitude of the departure.   

Although the Pritchet function and uniform magnitude distributions 
are widely used in the literature,
our analysis indicates that these assumptions require great caution:
when the actual counts follow a steeply rising luminosity function, 
incompleteness corrections based on a uniform magnitude distribution 
overestimate the number of real sources. 
We adopt the results from the artificial star tests with the Gaussian magnitude
distribution, since this more closely approximates the expected GCLF.  The
magnitude distribution of background sources is also a rising function, more
similar to the bright side of the Gaussian than to the uniform distribution.

The ratios of recovered to added artificial stars as a function of recovered
magnitude were calculated,
and a modified Fermi function was fitted to the completeness fractions in three annular
regions: from 0 to 33\arcsec, from 33\arcsec to 1\farcm1, and beyond 1\farcm1 
(red circles in Figure~\ref{fig_xy_nGC}). 
The values of $m_0$ found for these three regions are 29.30, 29.35, 
and 29.33 mag, respectively.
After applying the completeness corrections, the number of GC candidates brighter
than our adopted limit $\Iacs<29.32$ increases from 8212 to $8710{\pm}100$, or 6\%. 
Had we used the completeness estimates based on the uniform magnitude distribution, 
the increase would instead have been 14\%,
overestimating the final GC number by~7.5\%.

\begin{figure*}[!ht]
\begin{center}
\includegraphics[angle=0,width=0.48\textwidth]{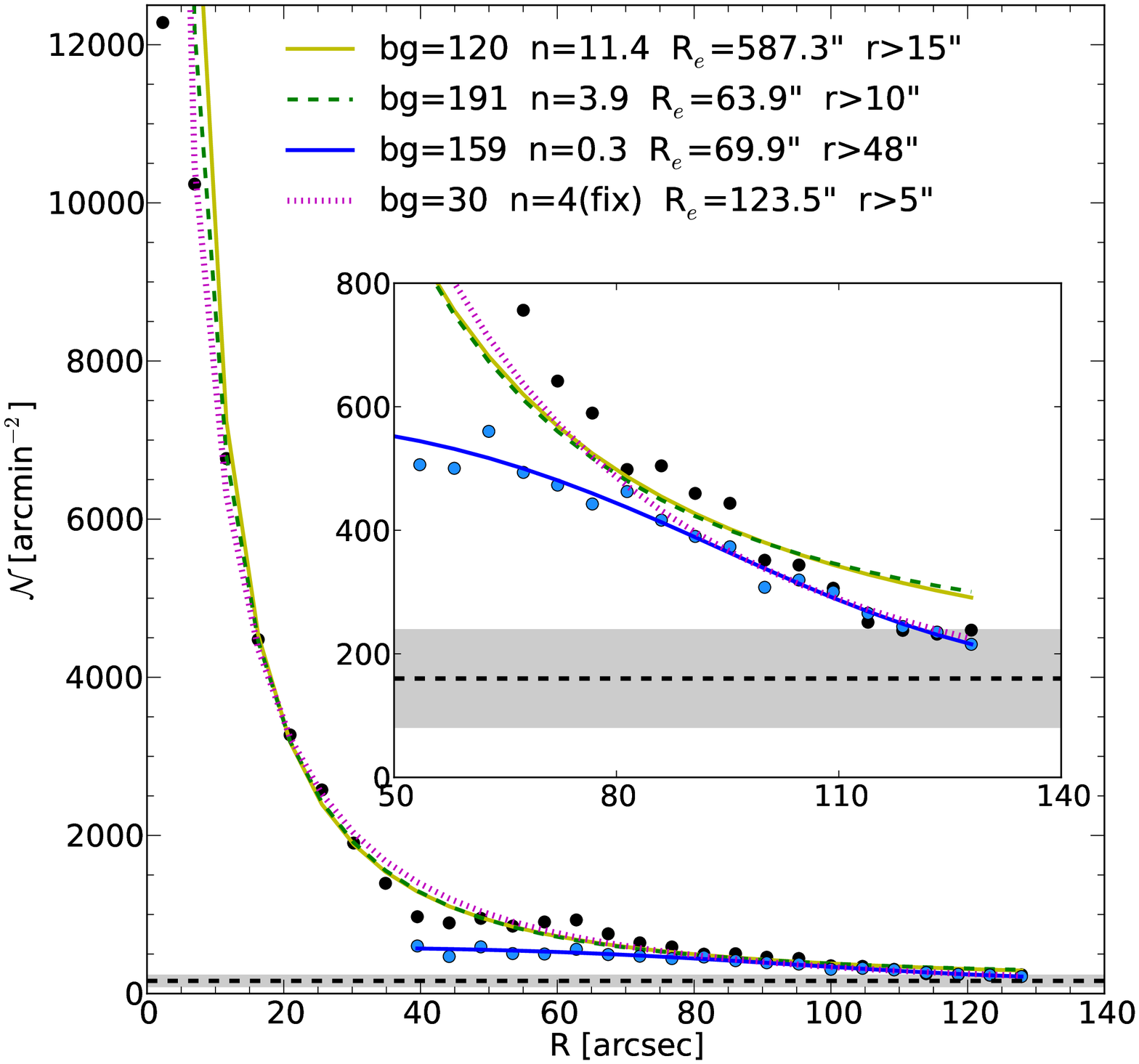}
\includegraphics[angle=0,width=0.48\textwidth]{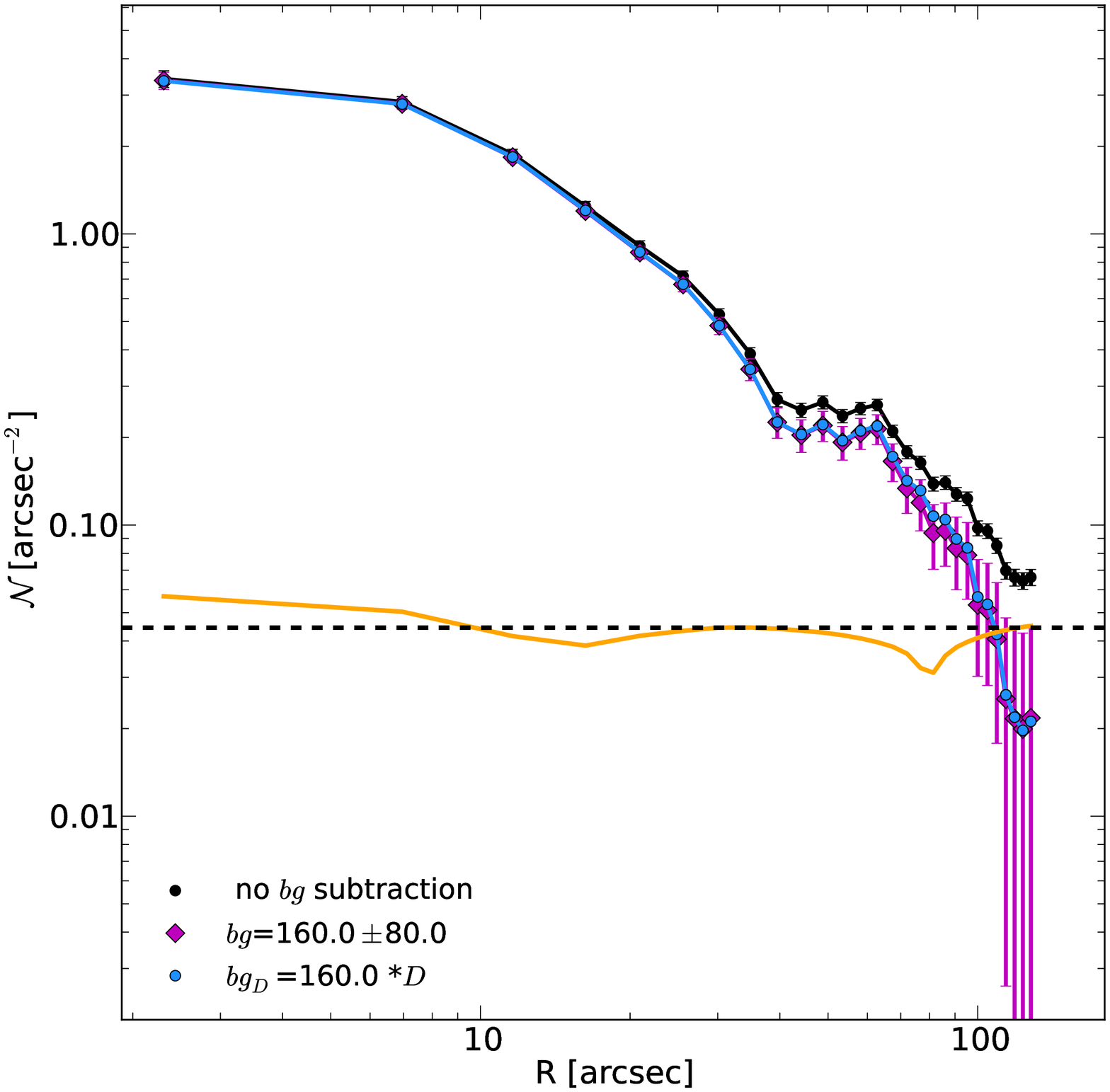}
\end{center}
\caption{Radial distribution of the surface number density of GC candidates.
{\it Left panel:} number of GC candidates per arcmin$^2$ (black dots). 
The dotted magenta, dashed green, and solid yellow lines show fitted S\'ersic functions
for different radial domains with the background (\bg) as an additional free
parameter. The blue dots are similar to the black ones, but obtained after all the bright galaxies
have been masked; the blue line shows the corresponding S\'ersic fit.
The S\'ersic parameters of the various fits are shown at top. 
The dashed black line and gray shaded region indicate the final adopted
background and 1-$\sigma$ error: \bg\ = (160$\pm$ 80) arcmin$^{-2}$.  
{\it Inset:} zoom of the outer parts of all fits.  
{\it Right panel:}
logarithmic plot of the number of GC candidates per arcsec$^2$ without background
subtraction (black dots), after subtracting a constant density of 160~arcmin$^{-2}$
(magenta diamonds), and after subtracting $160{\,\times\,}D$ arcmin$^{-2}$ (blue dots),
where $D$ represents the expected profile of the background dilution over
this magnitude range as a result of the lens magnification (see text).
The dashed black line indicates a value of 160\,arcmin$^{-2}$, while the
solid orange line indicates $160{\,\times\,}D$~arcmin$^{-2}$.
\label{GCdens_radial}}
\end{figure*}

\subsection{Background Contamination}
\label{bgcont.sect}

To estimate the number of background contaminants \bg, we first calculate 
the number of GC candidates per unit area as a function of radius (Figure~\ref{GCdens_radial}).
Then, we fit to the surface number density profile, $\mathcal{N}(r)$, 
a S\'ersic function (S\'ersic 1968) plus the background \bg\ as a free parameter:
\begin{equation}
\mathcal{N}(r)=\mathcal{N}_e\exp\left\{-b_n \left[ {\left( \frac{r}{R_e}\right) }^{1/n}- 1\right] \right\} + bg,
\end{equation} 
\noindent
where $R_e$ is the effective radius that encloses half of the GC sample; $\mathcal{N}_e$ is $\mathcal{N}$ at $R_e$; $n$ is the S\'ersic index, which controls the shape of the profile;
and $b_n\approx1.9992n - 0.3271$ (Graham \& Driver 2005). 

We found that the fitted S\'ersic parameters and background were very 
sensitive to the radial range of the fit. 
This is partly because of deviations from a smooth profile, but also 
because the number density of sources 
continues to decrease as a function of radius, without reaching a constant level.
However, the mean background within our adopted magnitude limits 
must be in the range $0<\bg\lta240$ arcmin$^{-2}$ 
(the high value being the observed density in the outermost bins).
After trying various radial cuts, we found the most robust 
value of \bg\ resulted by masking the regions around all the 
bright galaxies (gray region in the right panel of Figure~\ref{fig_xy_nGC}, 
based on our luminosity model).
This removes the concentrations of point sources associated with individual galaxies;
the remaining objects follow a shallower,
smoother radial density profile, shown in the left panel of Figure~\ref{GCdens_radial}. 
The fitted background in this case was $\bg\approx160$ arcmin$^{-2}$,
which was within the broad range returned by the various fits
prior to the galaxy masking (see examples in left panel of Figure~\ref{GCdens_radial}).
We adopt a conservative uncertainty of $\pm80$~arcmin$^{-2}$,
where the error bar encompasses the unlikely case that all objects in 
the outermost radial bins are background objects.

We note that after masking the bright galaxies, the remaining objects (blue points
in the left panel of Figure~\ref{GCdens_radial}) may represent a smooth population of IGCs
in A1689. If we integrate the S\'ersic function for these putative IGCs over the
full area of the image (including masked regions, since IGCs would also be projected
onto the galaxies), we find that 50\% of the GC candidates are in this 
smooth component.  Of course, some of these objects will be associated with
galaxies, since the spatial distributions of GCs are often more extended than the
starlight.  We therefore consider 50\% to be an upper limit on the IGC fraction
within this central region.  More detailed modeling of the GC distributions of
individual galaxies, and wider coverage to trace the profile of the smooth component
to larger radii, would help to refine this estimate.

Our estimate of the background is based on the outer parts of the ACS/WFC image, but
in the case of A1689, it is necessary to make a radial-dependent correction for 
the effect of cluster lensing (Blakeslee 1999).
The gravitational field of A1689 affects the spatial and magnitude 
distributions of the background sources.
Following the formalism of Broadhurst et al.\ (1995), in the absence of lensing
we expect a power-law distribution of background sources
$N_{bg}(m)\ \approx N_0 10^{\beta{m}}$, where $N_0$ is a constant, and 
$\beta$ is the logarithmic slope of the counts.
The lensing magnifies the brightness of the sources by the position-dependent
magnification factor $A$, and thus shifts the source magnitudes brighter by $2.5\,\log(A)$.
It also increases the surface area by the same factor, and thus decreases the surface 
density.  As a result, the effect of the lens magnification can be approximated as
\begin{equation}
N^{\prime}_{bg}(m) = A^{-1}\,N_0\,10^{\beta(m+2.5\,\log{A})}=N_{bg}(m)\,A^{-2.5(0.4 - \beta)}\,,
\end{equation} 
where $N^{\prime}_{bg}$ is the observed (lensed) number density.
Generally, $\beta<0.4$, so the background counts over a fixed magnitude range
are decreased, or diluted, by a factor $D = A^{-2.5(0.4 - \beta)}$. The
counts can also be amplified ($D>1$) in regions where $A<1$
(see Broadhurst et al.\ 1995 for a detailed discussion).

The magnification $A$ depends on the distance and mass distribution of the lens, 
as well as on the distance to the source plane.  In general, it can be written as:
\begin{equation}
A=\frac{1}{\mid{(1-\kappa)^2 - \gamma^2}\mid},
\end{equation} 
where $\kappa$ and $\gamma$ are the convergence and shear, respectively, 
for a given source distance.
We use a non-parametric $\kappa$ map for A1689 (M.\,J.\ Jee et al.\ 2013, 
in~prep.; see also Jee et~al.\ 2007), and since we are calculating the 
number densities in circular annuli, we adopt the spherical approximation
$\gamma=\bar{\kappa}-\kappa$, where $\bar{\kappa}$ is the mean convergence interior
to radius $r$.  Taking $\beta=0.35$ (e.g., Ben\'itez et al.\ 2004), 
we finally derive the dilution factor $D$ as a function of radius;
the orange line in the right panel of Figure~\ref{GCdens_radial} represents the product of 
$D$ and the background at large radius.  The background level \bg\ has
been normalized to the outermost several bins; interior to this, the dilution is 
both negative and positive, depending on radius.

We note that Coe et al.\ (2010) also constructed a $\kappa$ map for
A1689 using their ``LensPerfect'' algorithm, and they reported the best-fitting 
S\'ersic model parameters for the radially averaged profile.
As a check, we derived a background source dilution profile using 
Coe et al.'s S\'ersic parameters;
the differences with respect the above analysis was less than 1\%.  

Finally, we use the radial surface density distribution to correct 
for the incomplete area coverage (accounting for masked regions and 
incomplete outer annuli) out to a projected radius of 400~kpc (130\arcsec).
This increases the sample of GC candidates to 10,596, and 
subtraction of the radial-dependent background contamination then gives a
final sample of $8417{\pm}1096$ GC candidates with $\Iacs<29.32$ and $r<400$~kpc.

\section{Total GC Number and Specific Frequency}
\label{N_Sn.sect}

\begin{figure}
\centering
\includegraphics[angle=0,width=0.45\textwidth]{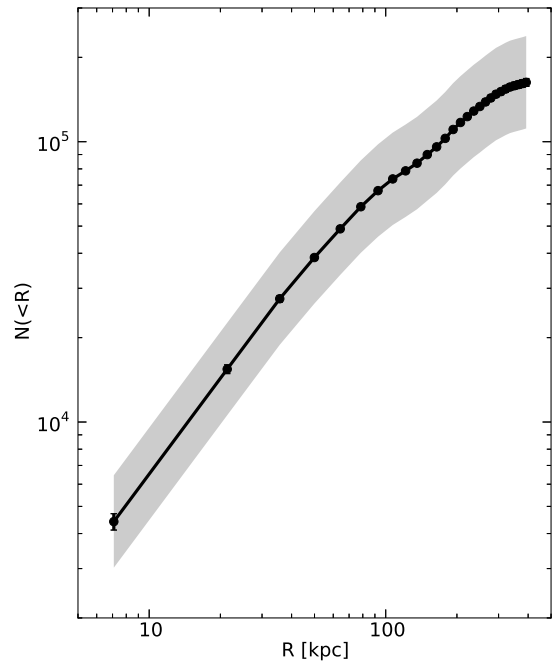}
\caption{Cumulative radial profile of the total number of GCs corrected for 
magnitude and area incompleteness and background contamination, then extrapolated
over the GCLF. 
The gray region shows the uncertainty due to the GCLF parameters. 
\label{Ncumulative}}
\end{figure}

After applying the corrections for incompleteness, partial area coverage, and
background contamination, we have ended up with a sample of 8417 GCs with
$m<29.32$ and within $r<400$ kpc of the central cD galaxy in A1689.
To estimate the total size of the GC population, we assume that the GCLF is similar
to those studied in massive ellipticals in more nearby clusters (e.g., Harris 2001;
Jord\'an et al.\ 2007; Harris et al.\ 2009; Peng et al.\ 2009; Villegas et al.\ 2010).  
Based on these works, the absolute \Iacs-band GCLF turnover should occur
at $M^{\rm TO}_{814} = -8.10{\pm}0.10$ AB~mag.
Including the distance modulus, Galactic extinction, and a $K$-correction in F814W of
0.03~mag (calculated for a GC spectrum at this redshift) implies an apparent
turnover magnitude of 31.71~mag. However, the lookback time to $z{=}0.183$ is
2.25~Gyr, and stellar population models (Bruzual \& Charlot 2003) imply that old,
metal-poor systems such as GCs would have been $\sim\,$0.15~mag more luminous at
this epoch. This is probably an upper limit because GC formation likely
occurred earlier in A1689 than in local clusters such as Virgo, Coma, or Fornax (see
the related discussion in Villegas et al.\ 2010).  We therefore adopt an apparent
turnover magnitude $m^{\rm TO}_{814} = 31.6\pm0.2$~AB and use a
Gaussian width $\sigma_{\rm LF}=1.4\pm0.1$~mag.

The extrapolated total population of GCs within $r<400$ kpc is then
$N_{\rm GC}^{\rm total}$=162,850${\pm}^{75450}_{51310}$. 
The main source of error comes from the uncertainty in the GCLF parameters; 
for instance, using $\sigma_{\rm LF}=1.3$ increases $N_{\rm GC}^{\rm total}$ by 30\%.
Figure~\ref{Ncumulative} shows the radial cumulative profile of 
$N_{\rm GC}$, including the uncertainty region.
For comparison, based on the Next Generation Virgo Survey (Ferrarese et al.\ 2012),
the total number of GCs within the same 400~kpc radius of M87 in the center of 
the nearby Virgo cluster is 26,400$\pm$3,200 (P.~Durrell et al., in preparation), 
a factor of six lower.

\begin{figure}
\centering
\includegraphics[angle=0,width=0.45\textwidth]{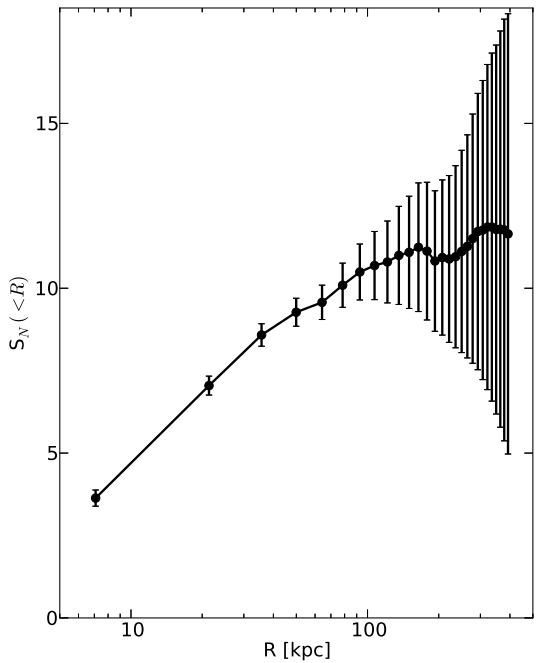}
\caption{Cumulative specific frequency $S_N$ as a function of radius; the error bars
  include statistical uncertainties in the number of GCs for the assumed (fixed)
  GCLF and the uncertainty in the galaxy light profile. \label{Scumulative}}
\end{figure}

The specific frequency, $S_N$, is defined as 
\begin{equation}
S_N \;=\; N_{\rm GC}\,10^{0.4(M_V + 15)}\,,
\end{equation}
where $M_V$ is the absolute magnitude of the galaxy in $V$ band, and both \ngc\ and
$M_V$ are measured over the same physical region. Since our data are in \Iacs, we need to apply a 
photometric transformation to obtain $M_V$. 
The absolute magnitude $M_V$ is derived as: 
\begin{equation}
M_V \;=\; I_{814} \,-\, \mM \,-\, A_{814} \,-\, K_{814} \,+\, (V{-}\Iacs)
\end{equation}
where \mM\ is the distance modulus, $A_{814}$ is the Galactic extinction,
$K_{814}$ is the $K$ correction, and $(V{-}\Iacs)$ is the rest-frame color.
We calculate $K_{814} = 0.11$ mag for the spectral energy
distribution of a giant elliptical at $z = 0.183$, and based on the
extensive compilation of $(V{-}I)$ galaxy colors by Tonry et al.\ (2001), 
we adopt $(V{-}\Iacs)=0.83$ (AB mag), in order to obtain $M_V$. 
Figure~\ref{Scumulative} shows the resulting cumulative $S_N$ as a function of radius
with the errors propagated from the GC counts and the galaxy luminosity.   
The uncertainty in the luminosity comes mainly from the assumed sky level
$\mu_{814}^{sky}=20.90{\pm}0.01$ mag\,arcsec$^{-2}$, causing the 
size of the $S_N$ error bars to increase strongly with radius.

\begin{figure*}
\centering
$\vcenter{\hbox{\includegraphics[angle=0,width=0.54\textwidth]{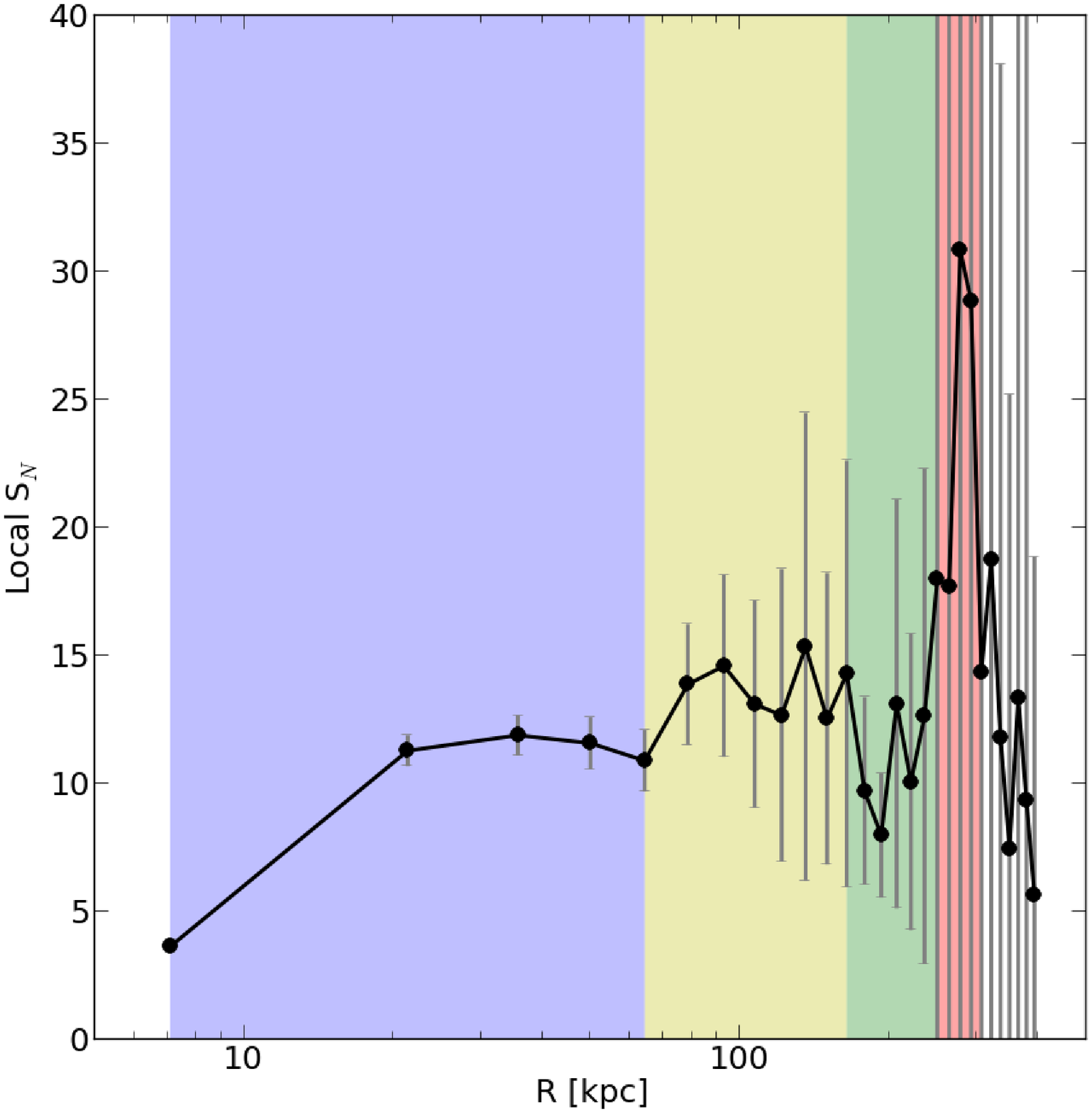}}}$
$\vcenter{\hbox{\includegraphics[angle=0,width=0.44\textwidth]{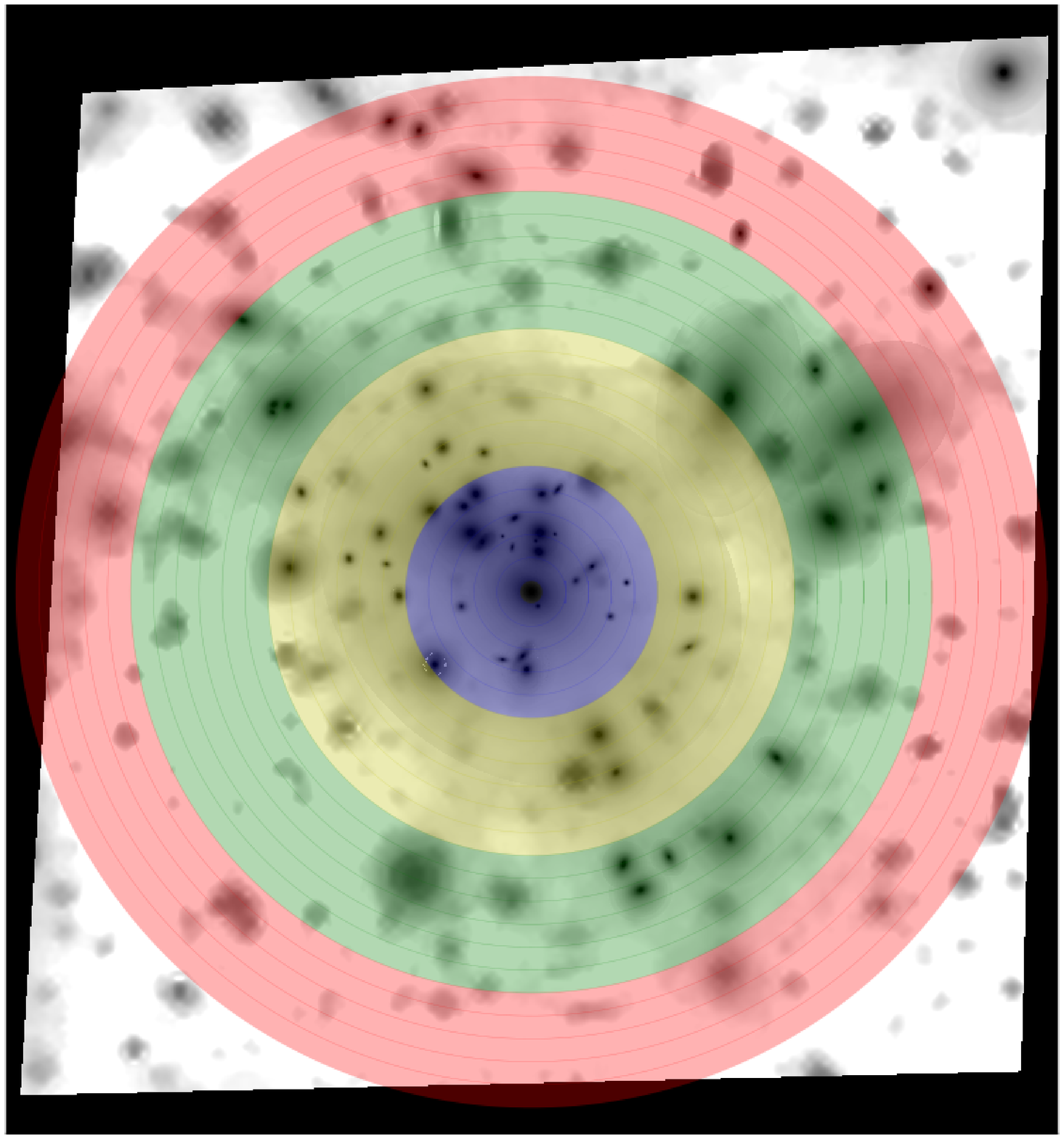}}}$
\caption{{\it Left panel:} local $S_N$ within $\sim5\arcsec$ annuli as a function of radius.
{\it Right panel:} galaxy light model. In both panels, the 
blue, yellow, green and red regions indicate radial ranges of
0-70, 70-180, 180-230 and 230-300 kpc, respectively. 
The ``dip'' in the local $S_N$ at $R{\,\approx\,}200$ kpc is caused by the grouping of bright
galaxies within the green annulus.\label{SnLocal_color}} 
\end{figure*}

For comparison to nearby galaxies, we need to consider how $S_N$ 
would evolve over a lookback time of 2.25~Gyr.
There is no significant star formation in the A1689 early-type galaxies 
that dominate this central field (Balogh et al.\ 2002), and
stellar population models that match the colors of nearby giant
ellipticals (e.g., 10~Gyr, solar metallicity models of Bruzual \& Charlot 2003)
indicate that they have passively faded by about 0.20~mag since $z=0.183$. 
Assuming negligible destruction of GCs in this time,
the passive galaxy evolution will cause $S_N$ to become 20\% higher at $z{\,=\,}0$
than at the observed epoch of A1689.
Thus, the global value of $S_N=11.7$ within 400~kpc would correspond to $S_N=14.0$
at z=0, after correcting for passive evolution, similar to the $S_N$ values observed for
the central cD galaxies in the nearby Virgo, Coma, and Hydra clusters 
(Tamura et al.\ 2006; Peng et al.\ 2008, 2011; Harris et al.\ 2009; Wehner et al.\ 2008). 
Of course, for a more exact comparison, the $S_N$ values should be
estimated on similar physical scales. 
Blakeslee (1999) measured $S_N$ within apertures of 40 and 65\,kpc for the cores of six 
rich Abell clusters; he reported $\langle{S_N(40)}\rangle=8.7$ and 
$\langle{S_N(65)}\rangle=9.2$ (with rms scatters of $\sim\,$2 in both cases). 
The corresponding values for A1689 at z=0 are $S_N$(40)=10.6$\pm$0.4 and
$S_N$(65)=11.5$\pm$0.5, near the high end of the observed range for nearby 
massive clusters.

In addition to the cumulative $S_N$, it is also worth considering the behavior of the
local $S_N$. 
We have noted that the surface density distribution of GC candidates is not completely
smooth, since they are preferentially located around the bright galaxies.
This clustering causes the ``bump'' at $\sim1\farcm1$ (200 kpc) in the GC 
radial density profile shown in Figure~\ref{GCdens_radial}.  This feature in the
GC density profile corresponds to a grouping of bright galaxies at this radius, as
highlighted within the green annulus in the right panel of Figure~\ref{SnLocal_color}.
However, since the GCs are not perfect tracers of the stellar light, and their spatial
concentrations are less sharply peaked than the galaxy profiles, an excess of
galaxies at a given location tends to \textit{decrease} the local value of $S_N$,
if the galaxies have a ``normal'' $S_N\approx4$.
Therefore, at the same radius of 200~kpc where there is a ``bump'' in the number
density of GCs, we actually find a ``dip'' in the local value of $S_N$, as shown by
the green band in the left panel of Figure~\ref{SnLocal_color}.
A corresponding dip occurs near 200 kpc in the cumulative $S_N$ distribution in 
Figure~\ref{Scumulative}.

The scaling of \ngc\ in brightest cluster galaxies with the total underlying
mass within a common projected radius has been interpreted as a consequence of
early universal GC formation efficiency in dense regions.
Using the value 0.71$\pm$0.22 GCs per $10^9$\mo (Blakeslee 1999), our derived
$N_{\rm GC}^{\rm total}$ would predict a total mass of $\sim2.3\times{10}^{14}$\,\mo\
within 400 kpc in A1689.  However, this estimate depends on the GCs following
the same radial profile as the total matter distribution.
While observations indicate that their spatial distribution is more extended than 
the starlight, until now it has not been possible to test how they relate to the
dark matter distribution.  We explore these issues in the following section.


\section{Comparison of Mass Profiles}
\label{results.sect}

\begin{figure*}
\centering
\hspace{-0.3em}
\includegraphics[angle=0,height=0.48\textwidth]{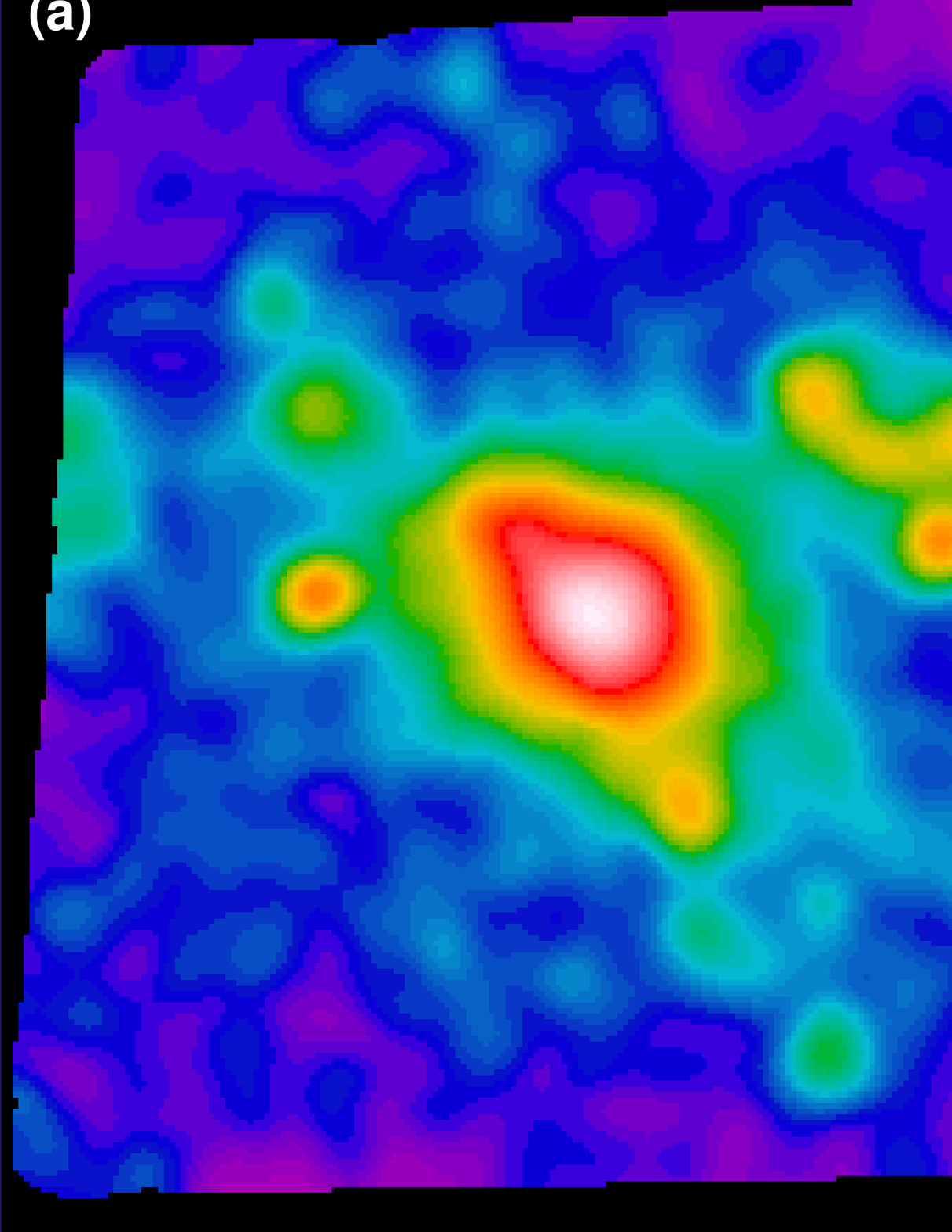}
\hspace{1.7em}
\includegraphics[angle=0,height=0.485\textwidth]{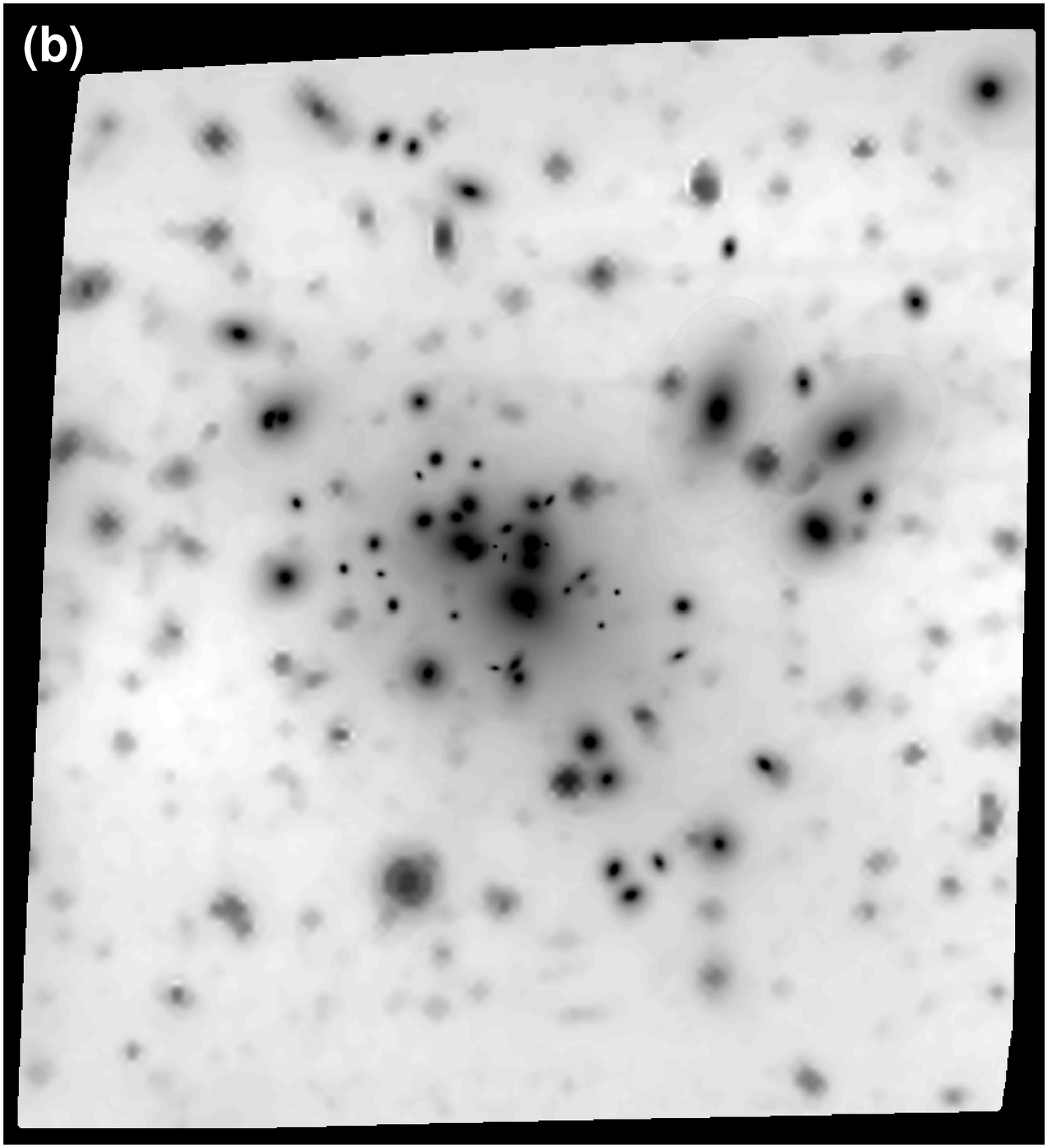}
\vspace{2mm}
\hspace{-3em}
\includegraphics[angle=0,height=0.48\textwidth]{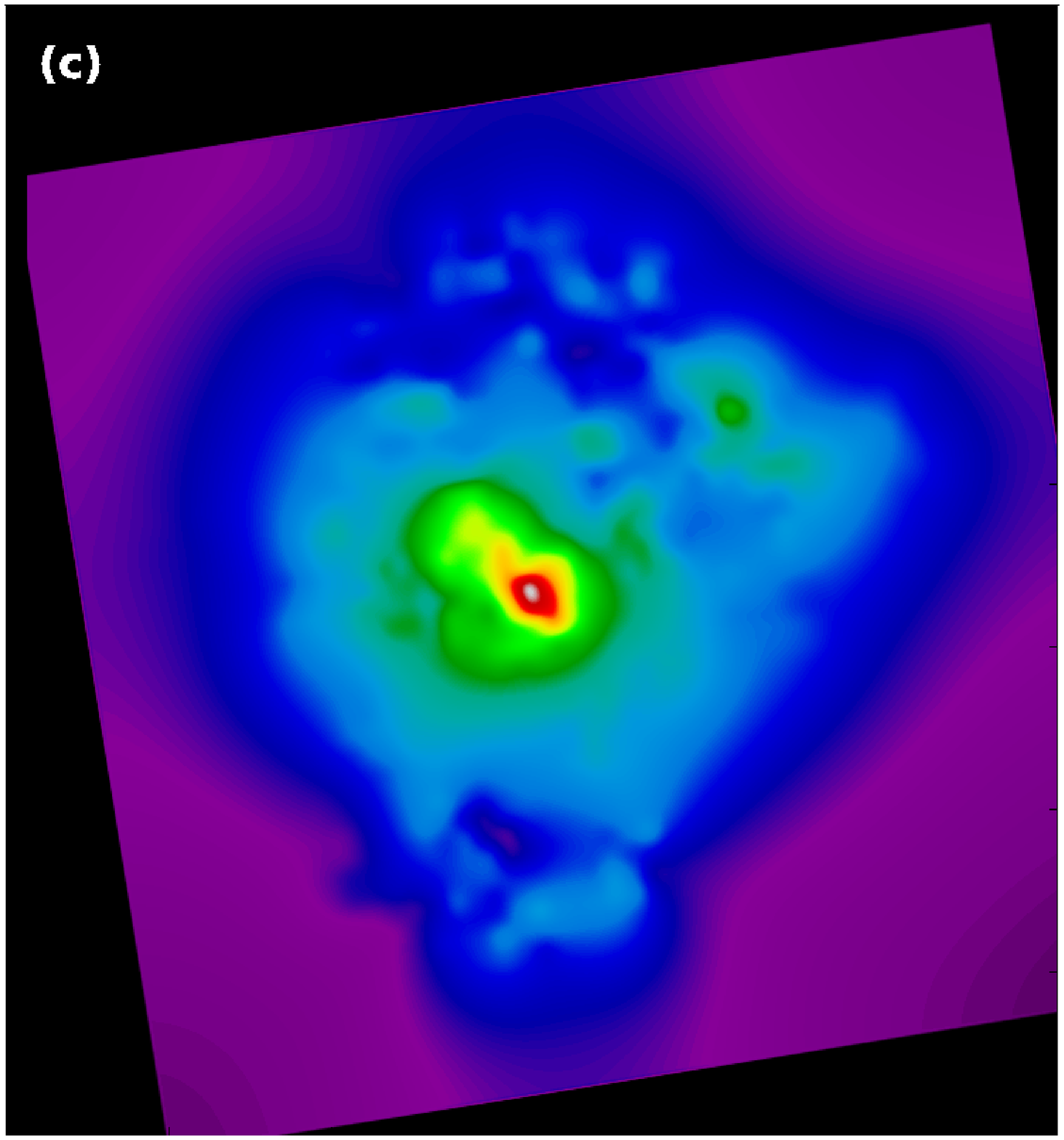}
\includegraphics[angle=0,height=0.48\textwidth]{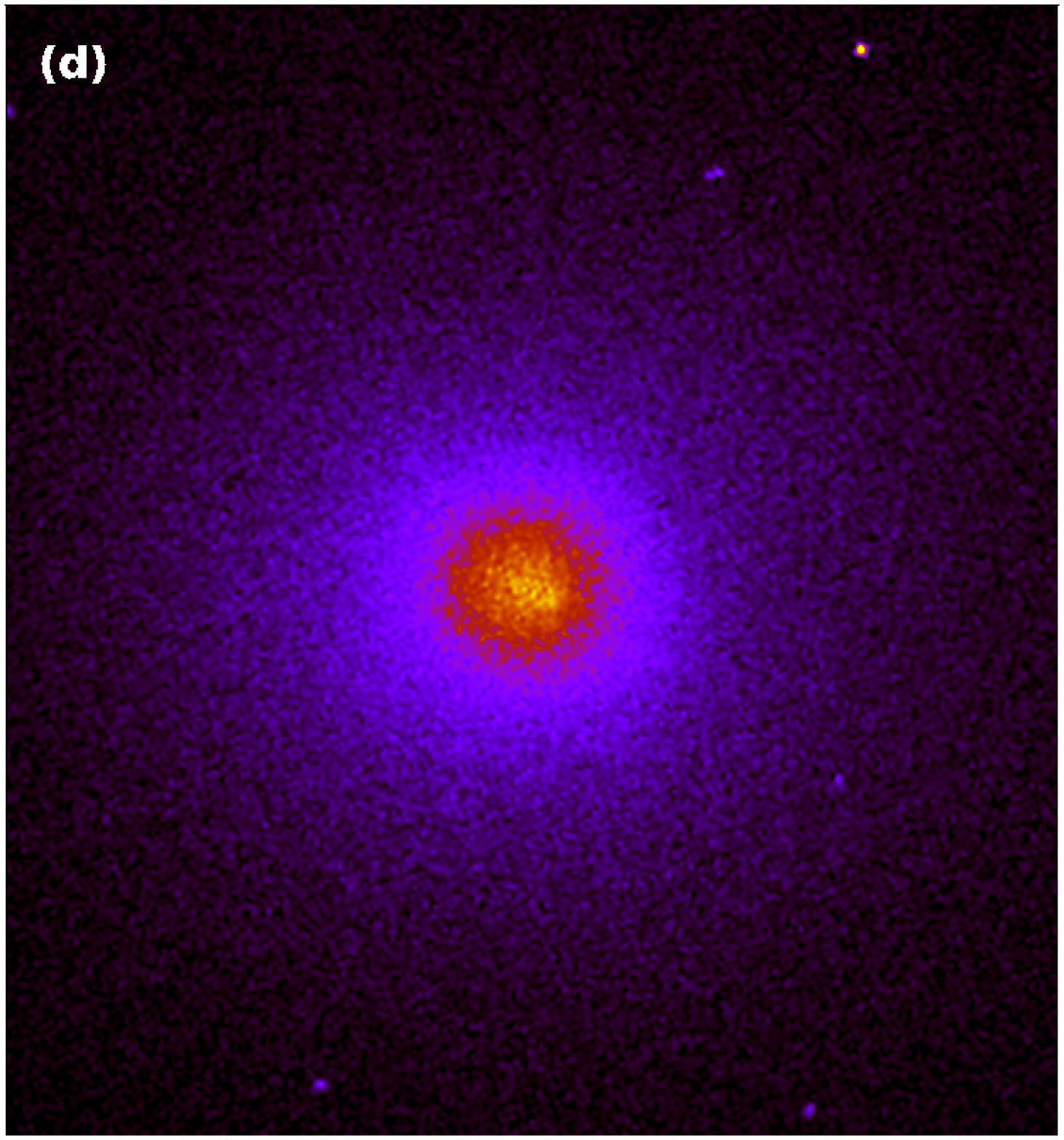}
\caption{{$(a)$}~Surface number density of GC candidates, smoothed with a Gaussian
of FWHM = 10\arcsec\ (300~pix);
{$(b)$}~surface brightness distribution from 59 galaxies with isophotal models
plus the SExtractor background map; 
{$(c)$}~lensing-derived total surface mass density (includes baryonic 
and non-baryonic components); 
{$(d)$}~X-ray surface brightness distribution (from {\it Chandra} archive). 
\label{GC_stars_ICG_lens.ima}}
\end{figure*}

With the goal of testing the existence of a universal GC formation efficiency in
an extreme system such as A1689, we compare the amounts of mass in this central
field in the form of GCs, stars, hot intracluster gas, 
and total mass (including dark matter). 
Figure~\ref{GC_stars_ICG_lens.ima} provides a 2-D visual comparison of the 
number density of GCs, the galaxy luminosity model, the lensing-derived mass
distribution, and the X-ray emission map.  The symmetry and smoothness of
the X-ray gas stand out; this is not a resolution effect (as evidenced by the compactness
of the X-ray point sources in the image). 
To make quantitative comparisons, we now derive the projected radial 
mass distributions for each component.

For the GCs, the total number within 400~kpc calculated in the preceding section can 
be converted to a mass by assuming a mean individual GC mass of
$\langle$\M$_{\rm GC}\rangle = 2.4{\,\times\,}10^5$\mo\  
(McLaughlin 1999; Blakeslee 1999).
The estimated total mass in GCs within this radius is therefore 
$\M_{\rm GC}^{\rm total}$=3.9$\times{10}^{10}$\mo.
For perspective, this is 60--80\% of the total stellar mass
of the Milky Way galaxy (Flynn et al.\ 2006; McMillan 2011).

To calculate the stellar mass \M$_{\star}$ we assume a stellar mass-to-light ratio 
$\M_{\star}/L_V{\,=\,}4$, based on solar metallicity Bruzual \& Charlot (2003)
models with a Salpeter initial mass function (IMF).  
Of course, the assumption of a constant \M$_{\star}/L_V$ at all radii is 
a first-order approximation.
Moreover, the $\M_{\star}/L_V$, and thus the derived mass, can vary by $\sim\,$30\%,
depending on the IMF, but the choice of Salpeter is reasonable for early-type
galaxies (e.g., van~Dokkum \& Conroy 2010; Conroy \& van~Dokkum 2012).
Finally, using $M_{V, \odot}=4.81$, we obtain \M$_{\star}^{total}$=4.7$\times{10}^{12}$\mo.

The X-ray gas mass was estimated from the 3-D gas density profile $\rho_g(r)$
constructed by Lemze et al.\ (2008) based on {\it Chandra} X-ray data.
We found that the published $\rho_g(r)$ is well fitted by a function of the form:
\begin{equation}
\rho_g(r) \;=\; \frac{\rho_0}{[1 \,+\, (r/r_0)^\alpha]^\beta}\,,
\end{equation}
with best-fit values $\rho_0=2.1\times{10}^{-25}$g\,cm$^{-3}$, $r_0=321$ kpc, 
$\alpha=0.58$, and $\beta=5.6$
(giving $r^{\,-\alpha\beta} \sim r^{\,-3.2}$ at large $r$).
These parameters are only used for interpolating the X-ray data points.
We then integrate this function along the line of sight $\ell$ to obtain the projected
gas mass surface density $\zeta$ as:
\begin{eqnarray}
\zeta \;=\; 2\int_{0}^{1\,{\rm Mpc}}\rho(\sqrt{R^2 + \ell^2})d\ell\,,
\end{eqnarray}
where R is the projected radius. 
This gives an X-ray gas mass within 400 kpc of 
$\M_{\rm X-ray}^{\rm total}=3.6\times{10}^{13}$\mo. 
The baryonic mass comprises field stars, GCs, and intracluster gas.
Adding all the components, we obtain \M$_{\rm baryon}\approx4.1\times{10}^{13}$\mo.

The total mass, \M$_{\rm total}$, is estimated from the $\kappa$ map used
previously in Sec.\,\ref{bgcont.sect}.
It includes both baryonic and non-baryonic mass, 
but is dominated by the non-baryonic component (at least outside the central few kpc).
The convergence $\kappa$ is the mass surface density, 
normalized by the critical surface density (i.e., $\kappa{\,=\,}\Sigma$/$\Sigma_{crit}$), where 
\begin{equation}
\Sigma_{crit}=\frac{c^2 D_s}{4\pi{G}\,D_LD_{Ls}}\,; 
\end{equation}
$c$ is the speed of light; $D_L$, $D_s$, and $D_{Ls}$ are the distances to the lens (A1689), 
to a reference source (assumed $z{\,=\,}3$), and from the lens to the source, respectively. 
Integrating, we find \M$_{\rm total}$=6.4$\times{10}^{14}$\mo\ within 400~kpc.

\begin{figure*}
\centering
\includegraphics[angle=0,width=0.98\textwidth]{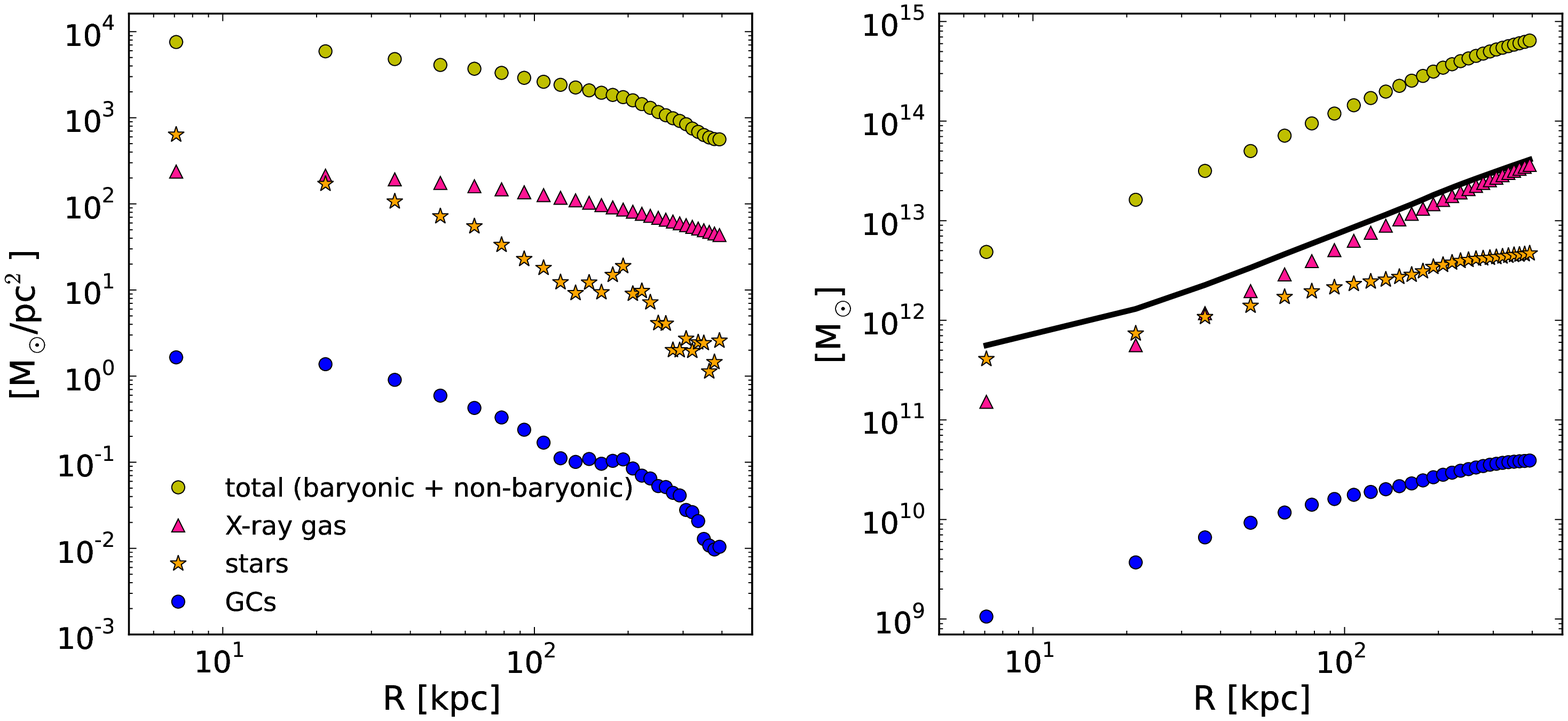}
\caption{{\it Left:} radial mass density profile for each component. 
{\it Right:} cumulative mass profiles versus radius. 
The thick black line indicates the cumulative profile for the baryonic mass,
$\M_{\rm baryon} = \M_{\rm X-ray}+\M_{\star}$. \label{massProfiles}} 
\end{figure*}

Figure~\ref{massProfiles} presents the radial mass density and cumulative mass
profiles for each of the above components. The stars and GCs are strongly
concentrated in and around the galaxies; the most prominent feature is the
``bump'' at $R\approx200$ kpc, discussed above. 
The hot X-ray emitting gas and total mass exhibit much smoother profiles.
The hot gas dominates the baryonic mass beyond the central $\sim30$ kpc. 

\begin{figure*}
\centering
\includegraphics[angle=0,width=1.0\textwidth]{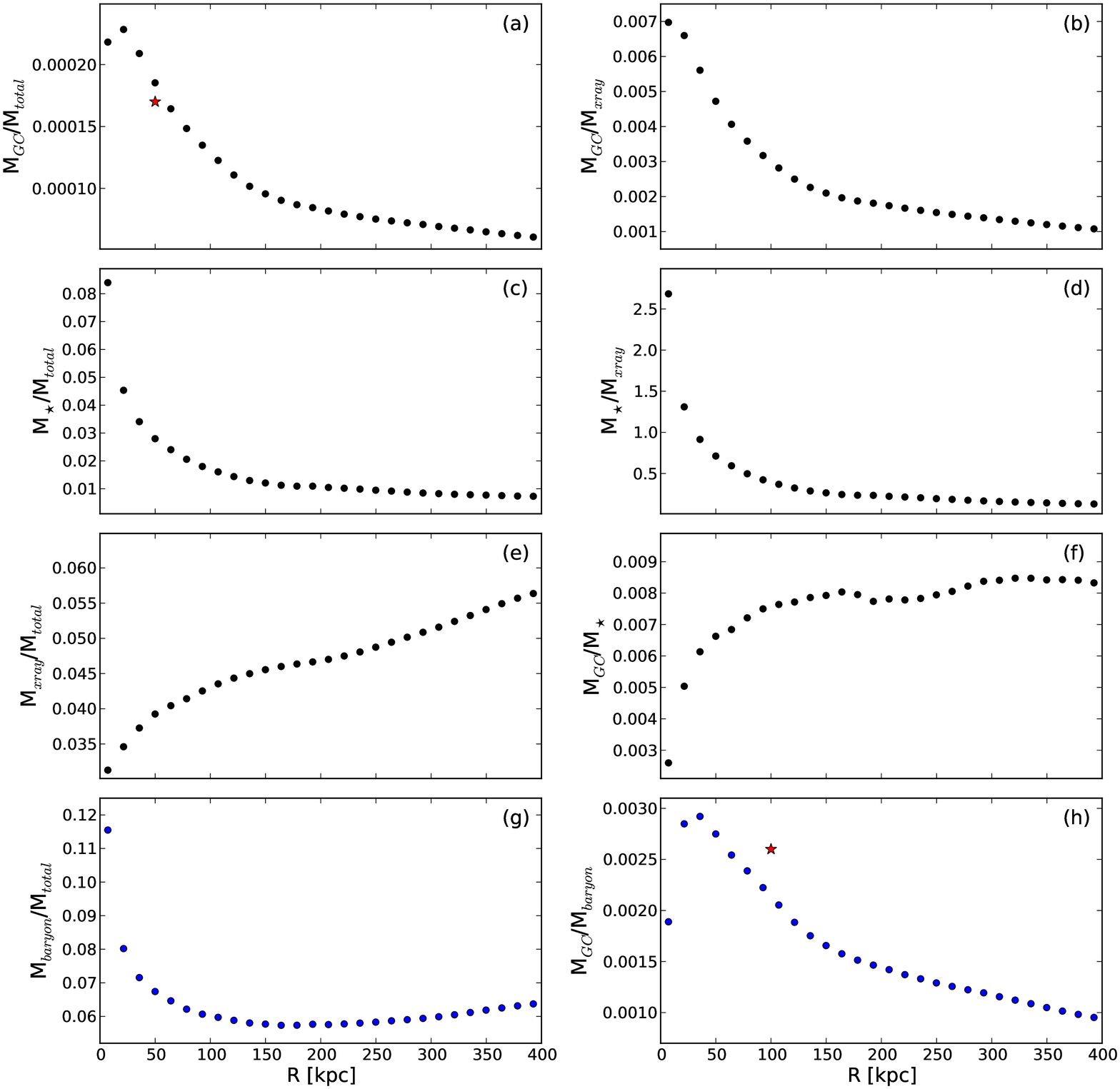}
\caption{Ratios of the cumulative distributions for various mass components in A1689,
based on the curves plotted in Fig.\,\ref{massProfiles}. 
The red stars in $(a)$ and $(h)$ represent the values
$\epsilon_t= 1.7{\times}10^{-4}$ from Blakeslee (1999) and
$\epsilon_b=0.0026$ from McLaughlin (1999), respectively.  
See text for discussion.
Note that the GC stellar mass fraction $\M_{\rm GC}/\M_{\star}$ shown in panel~$(f)$ is 
equivalent to $\smass/100$, where \smass\ is
the ``specific mass'' parameter used in some studies (e.g., Peng et al.\ 2008).
\label{ratios_plots}}
\end{figure*}

To investigate the radial run of the GC mass fraction relative to the other mass
components, we plot in Figure~\ref{ratios_plots} the mass ratios for all
possible combinations of the components in Figure~\ref{massProfiles}.
The stellar mass fraction in GCs ($\M_{GC}/\M_\star$, panel $f$) increases by a factor of
$\sim\,$2.5 within the central 80~kpc, then levels off, consistent with the
$S_N$ profile shown previously.  However, compared to either the X-ray gas or
total matter, the GC mass fraction decreases as a function of radius (panels $b$
and $a$, respectively); the mass fraction in stars (panels $d$ and $c$) shows a
similar, even steeper, decline.  
As found in previous studies, the gas mass fraction (panel $e$) appears to increase
monotonically with radius. 
Interestingly, the baryonic mass fraction (ratio of baryons to total mass, panel $g$)
reaches a minimum near 150~kpc, because the stellar mass is more concentrated than the
dark matter, but the gas is more extended.  This quantity has been studied extensively in
galaxy clusters as an approximation to the baryonic mass fraction of the universe
($\M_b/\M_{DM}\sim\Omega_{\rm b}$/$\Omega_{\rm DM}$).
For instance, Lin et al.\ (2012) measured baryon fractions for 94 galaxy clusters over a
wide range redshift; our value for A1689 in Figure~\ref{ratios_plots} agrees well with the
results of this recent study.

McLaughlin (1999) derived an average efficiency of GC formation per baryonic
mass, i.e., $\epsilon_b=$\M$_{\rm GC}$/\M$_{\rm baryon}$, of 0.0026 within
$r<100$ kpc. Comparing Mclaughlin's value with our estimate for the same radial
range (Figure~\ref{ratios_plots}$h$), we find $\epsilon_b$=0.0021, which
is within the scatter of the values found by McLaughlin.  
Meanwhile, Blakeslee (1999) reported a mean efficiency per total mass of $\epsilon_t=$\M$_{\rm GC}$/\M$_{\rm total} = 1.7{\times}10^{-4}$ 
within $r<50$ kpc;
within this radius, we obtain $\epsilon_t$=1.8$\times$10$^{-4}$
(Figure~\ref{ratios_plots}$a$), in close agreement with the expected value.
However, although the mass ratios within these relatively small radii are
remarkably consistent with the ``universal'' efficiencies previously proposed in
the literature, the global values are not.  Within 400 kpc, we find
$\epsilon_b^{A1689}$=0.00095 and $\epsilon_t^{A1689}$=6.1$\times$10$^{-5}$, both
a factor of 3 lower than the values cited above.  
After converting to our assumed mean GC mass, the results of Spitler \& Forbes (2009)
 imply $\epsilon_t=4.2\times10^{-5}$, about 30\% lower than our value within 400~kpc but
 consistent within the errors; however, it not clear what radius to use in this case.
Thus, we emphasize again the importance of comparing such ratios within 
the same physical radii. 

\section{Discussion and Conclusions}
\label{discussion.sect}

Deep broadband imaging with \hst/ACS ($\gtrsim\,$90\% complete to $I_{814}=29$) has
revealed an extremely rich GC system in the center of the massive lensing cluster A1689. 
The estimated total population of 162,850$^{+75,450}_{-51,310}$ GCs within a
projected radius of 400~kpc represents the largest system of GCs studied to date,
six times the number within the same radius in the Virgo cluster.
The large error~bars are due to the uncertainty in the GCLF parameters.  Although
the Gaussian form of the GCLF is well calibrated for giant ellipticals in rich clusters,
even with 20.9~hrs of integration, our data fall 1.6$\,\sigma$ short of the
GCLF turnover; we therefore sample only 10\% of the GCs brighter than the turnover
(or 5\% of the total population, assuming a symmetric GCLF).  Thus, the large
extrapolation yields a sizable systematic uncertainty.  Nevertheless, this remains the
largest GC system yet discovered, with at least a factor of two more than the next most
populous systems, including Coma and A3558 (Peng et al.\ 2011; Barber~DeGraaff 2011).

Our analysis accounts for the effects of Eddington bias, gravitational magnification of
the background surface density, redshifting of the bandpass ($K$~corrections), and
passive evolution of the GCLF.  Although it is possible that there has been some
evolution in the shape of the GCLF since $z=0.18$, this would mainly occur by the destruction
of low-mass GCs (e.g., Jord\'an et al.\ 2007; McLaughlin \& Fall 2008) with little
effect for the masses of objects near the GCLF peak or brighter. Our assumption of a
symmetric GCLF would therefore likely underestimate the population at $z=0.18$, but the
additional low-mass GCs would have little effect on our total mass estimates.

The spatial distribution of GC candidates in the center of A1689 is not completely
smooth; there are obvious concentrations around the cluster galaxies.  However, by
masking the bright and intermediate-luminosity galaxies, we can trace an
apparently smooth component, which comprises half of the total population when
integrated over the field (via the best-fit S\'ersic model).  If we identify these
objects as belonging to a possible intracluster population of GCs, then there may be as
many as $\sim\,$80,000 such IGCs within the central 400~kpc in A1689.  We consider this
an upper limit, since some of these IGC candidates are undoubtedly bound to individual
galaxies. We plan to investigate this in more detail by modeling the GC distributions
around individual galaxies.  Imaging to a similar depth at larger radii from the cluster
center would also help in constraining the IGC population.

The cumulative $S_N$ increases from a value near $5$ within 10~kpc to $10{\pm}0.5$ 
(not including systematic uncertainty from the GCLF) within 70~kpc. 
Although the uncertainties in $S_N$
become large beyond 100 kpc, the profile appears to flatten, and the value within 150~kpc
is $S_N=11.1\pm2.0$.  There is a clear dip in the cumulative $S_N$ around 200~kpc, before it
rises again to $\sim12$ within 300~kpc.  The dip at 200 kpc occurs despite a local
increase in the GC number density at the same radius; it is caused by a subgrouping of
several bright galaxies which appear to have a more normal $S_N < 10$ (as was shown by
comparing the green regions in Figure~\ref{SnLocal_color}).  Such galaxies contribute
relatively more to the denominator of $S_N$ than to the numerator.  This highlights the
fact that cannibalization of normal cluster galaxies by the central cD will tend to
decrease the $S_N$, rather than increase it; thus the high $S_N$ value must have been
imprinted in the cluster core at early times.
We have also noted that passive evolution of the galaxy luminosity since $z{\,=\,}0.18$ would cause
the observed $S_N$ to increase by 20\% at $z{\,=\,}0$, but even with this effect, the high
$S_N$ in A1689 would not be anomalous among cD galaxies in local clusters.

Remarkably, the mass in GCs within 400~kpc of the center of A1689 is equivalent to
60-80\% of the total stellar mass of the Milky Way.  Integrated out to the virial radius
of $\sim3$~Mpc, the total mass in GCs is likely twice that of the stars in our Galaxy.  
Of course, this is a small fraction of the total mass in A1689.
We have examined the mass profile of the GCs as a function of radius and compared
it with the mass profiles of the stellar light, hot intracluster gas, and total
lensing-derived matter content within this central field.
The mass profile of the GCs is somewhat more extended than the stellar light, but more
concentrated than the hot gas or dark matter.  If the mass fraction is viewed as a GC
formation efficiency, then the efficiency (in terms of either baryonic or
total mass) decreases as a function of radius, and there is no ``universal'' value.  

On the other hand, when compared within the same physical radii, the GC mass fractions with
respect to the total and baryonic masses agree with the values found in samples of
nearby clusters, all of which have masses lower than A1689.
This suggests the possibility of a universal GC formation profile within galaxy clusters.
In contrast,
Lagan\'a \etal\ (2011) estimated the stellar, intracluster gas, and total masses within
$r_{500}$ for 19 galaxy clusters, and found a decrease in the stellar mass fraction with
increasing total mass of the system.  That is, more massive clusters had lower
overall star formation efficiencies.  Taken together, these results are consistent with
the view that the high $S_N$ values in cD galaxies are a consequence of ``missing''
stellar light in more massive clusters (Blakeslee 1997), rather than an excess in the
number of globular clusters.

Finally, we note that in a recent study, Su\'arez-Madrigal et al.\ (2012) modeled the
influence of the dark matter halo on molecular clouds at different locations within it.
They found that the star formation efficiency of the clouds depends on the ambient
density, and thus decreases with distance from the halo center, in qualitative agreement
with our finding that GC formation efficiency decreases with radius.  If the mass density
profiles in galaxy clusters are approximately universal (e.g., Navarro et al.\ 1997), then
it would also make sense that the GC formation efficiency profile may follow a universal
form.  Further studies of the GC, baryonic, and total mass profiles in galaxy clusters are
needed to test this intriguing possibility.

\acknowledgments
Support for program GO-11710 was provided in part through a grant from the Space
Telescope Science Institute, which is operated by the Association of
Universities for Research in Astronomy, Inc., under NASA contract NAS5-26555.
We thank Dan Magee for help with the CTE correction software. JPB acknowledges helpful
conversation with James Bullock. K.A.A-M acknowledges the support of CONACyT (M\'exico).  
This research has made use of the NASA/IPAC Extragalactic Database (NED), 
which is operated by the Jet Propulsion Laboratory, California Institute of
Technology, under contract with NASA.

\noindent
{\it Facilities:} \facility{HST (ACS/WFC)}.



\end{document}